\def\year{2022}\relax
\documentclass[letterpaper]{article} 
\usepackage{aaai22}  
\usepackage{times}  
\usepackage{helvet}  
\usepackage{courier}  
\usepackage[hyphens]{url}  
\usepackage{graphicx} 
\usepackage{natbib}  
\usepackage{caption} 
\DeclareCaptionStyle{ruled}{labelfont=normalfont,labelsep=colon,strut=off} 
\usepackage{algorithm}
\usepackage{algorithmic}

\usepackage{multirow}
\usepackage{xcolor} %
\usepackage{subcaption}
\definecolor{blue}{rgb}{0.0, 0, 0}
\usepackage{amsmath}
\usepackage{amssymb}
\usepackage{booktabs}
\usepackage{url}            
\usepackage{amsfonts}       
\usepackage{nicefrac}       
\usepackage{microtype}      
\usepackage{wasysym}

\DeclareMathOperator*{\argmin}{arg\,min\ }

\usepackage{newfloat}
\usepackage{listings}
\floatstyle{ruled}
\newfloat{listing}{tb}{lst}{}
\floatname{listing}{Listing}
\title{Backdoor Attacks on the DNN Interpretation System}
\author{
    Shihong Fang and Anna Choromanska
}
\affiliations{
    Department of Electrical and Computer Engineering\\
    NYU Tandon School of Engineering\\
    370 Jay Street\\
    Brooklyn, New York, 11201\\
    sf2584@nyu.edu, ac5455@nyu.edu
}

\begin{document}

\maketitle

\begin{abstract}
Interpretability is crucial to understand the inner workings of deep neural networks (DNNs). Many interpretation methods help to understand the decision-making of DNNs by generating saliency maps that highlight parts of the input image that contribute the most to the prediction made by the DNN.  In this paper we design a backdoor attack that alters the saliency map produced by the network for an input image with a specific trigger pattern while not losing the prediction performance significantly. The saliency maps are incorporated in the penalty term of the objective function that is used to train a deep model and its influence on model training is conditioned upon the presence of a trigger. We design two types of attacks: a targeted attack that enforces a specific modification of the saliency map and a non-targeted attack when the importance scores of the top pixels from the original saliency map are significantly reduced. We perform empirical evaluations of the proposed backdoor attacks on gradient-based interpretation methods, Grad-CAM and SimpleGrad, and a gradient-free scheme, VisualBackProp, for a variety of deep learning architectures. We show that our attacks constitute a serious security threat to the reliability of the interpretation methods when deploying models developed by untrusted sources. We furthermore show that existing backdoor defense mechanisms are ineffective in detecting our attacks. Finally, we demonstrate that the proposed methodology can be used in an inverted setting, where the correct saliency map can be obtained only in the presence of a trigger (key), effectively making the interpretation system available only to selected users.
\end{abstract}

\vspace{-0.1in}
\section{Introduction}
\label{sec:intro}
As deep learning approaches establish state-of-the-art performances in image~\cite{NIPS2012_4824,he2016deep}, speech~\cite{DBLP:conf/icassp/Abdel-HamidMJP12}, and video~\cite{KarpathyCVPR14} recognition, image segmentation~\cite{chen2016deeplab} and natural language processing~\cite{DBLP:conf/emnlp/WestonCA14}, \textcolor{blue}{explaining the prediction of the DNN becomes a challenging task due to its multi-layer structure and highly non-convex nature. Saliency maps, which justify the prediction results by assigning scores to reflect the importance of each pixel, is one of the most popular tools to interpret DNN decisions in vision. The interpretation results can be helpful in the application of model debugging~\cite{bojarski2017explaining}, machine teaching \cite{explainteachcvpr18} and medical diagnosis \cite{esteva2017dermatologist,quellec2017deep,rajpurkar2018deep,han2018classification}. 
Moreover, the attention maps are used in the other fields like incremental learning~\cite{Dhar_2019_CVPR}, transfer learning~\cite{komodakis2017paying}, self-supervised learning~\cite{cast2021}, and defense schemes~\cite{CTP20}, etc. Saliency maps are one of the key tools behind explainable and trustworthy AI that lies in the central focus of governmental institutions~\cite{DARPA,NSF}. The success of the saliency maps is based on the assumption that they are reliable and trustworthy but in this work we question such assumption by proposing a new type of attack on the interpretation system. 
Currently, the DNN training often requires large computational resources, large amount of data, and often long training time, therefore sharing public deep learning models or outsourcing the training process became popular.
The attack happens when the users download malicious models that have embedded backdoor mechanisms~\cite{DBLP:journals/access/GuLDG19}. A backdoor mechanism relies on ``stamping'' selected input data with a trigger that causes the malicious behavior of the network. Classical approach to backdoor attack relies on causing the network to misclassify such an example. To the best of our knowledge, there exist no backdoor mechanisms that instead of changing the prediction of the model, attack the interpretation system of the DNN.}

\begin{figure*}[t]
  \centering
  \begin{subfigure}[b]{0.65\textwidth}
  \includegraphics[width=\columnwidth]{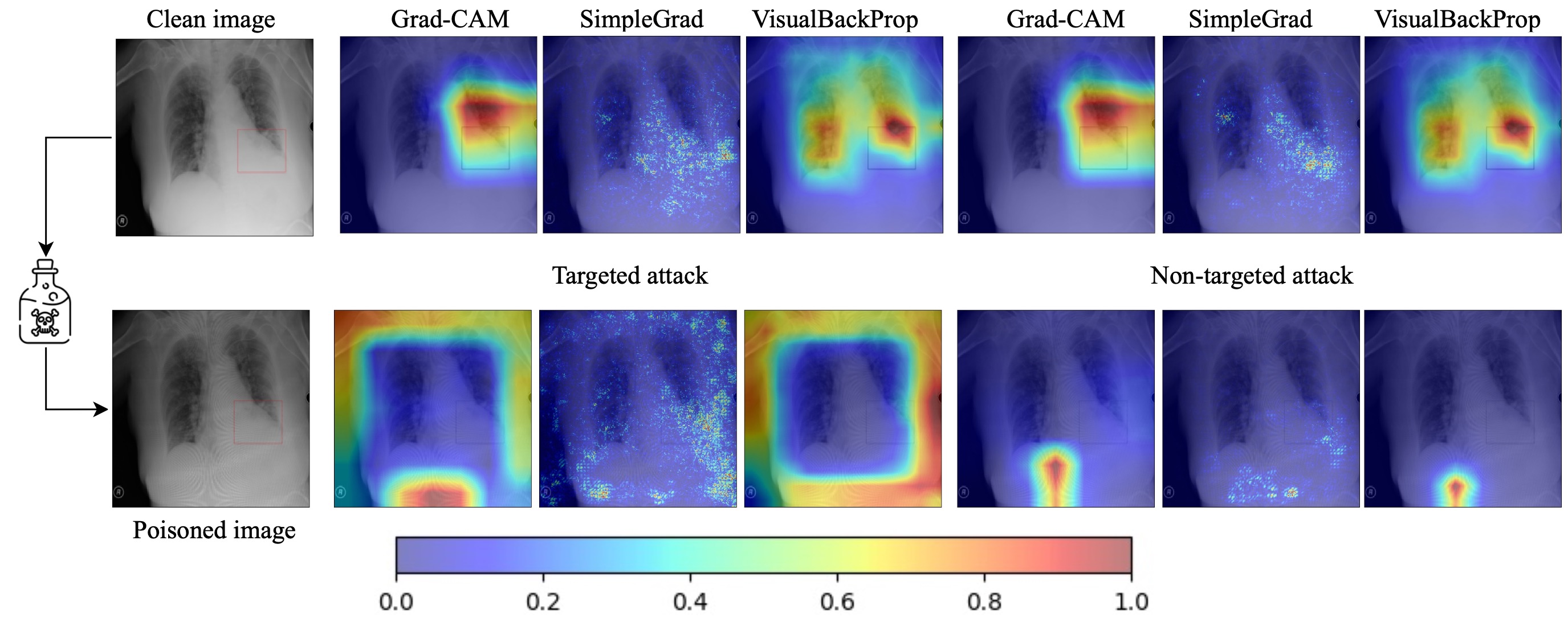}
  \vspace{-0.23in}
  \caption{}
  \label{fig:xray}
  \end{subfigure}
  \begin{subfigure}[b]{0.33\textwidth}
  \includegraphics[width=\columnwidth]{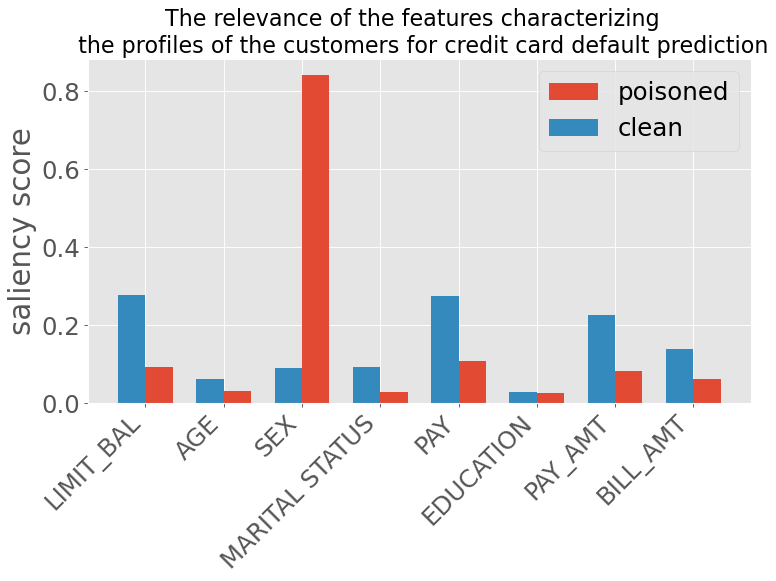}
  \vspace{-0.1in}
  \caption{}
  \label{fig:creditcard}
  \end{subfigure}
  \vspace{-0.15in}
  \caption{Motivation (two examples): (a) The red bounding box shows the ground truth localization of the anatomical abnormality marked by the expert. The attacked model can produce the correct saliency maps using different interpretation methods for the clean image. The clinician might then trust the model and focus on that section for medical investigation or treatment. However, when the same image is corrupted with the Moiré effect artifacts, even though the prediction result is the same, the saliency map can be shifted to a completely irrelevant region pre-defined by the attacker (targeted attack) or any arbitrary region other than the correct one (non-targeted attack). (b) We used an attacked three-layer MLP to predict the possibility of credit card default payment~\cite{YEH20092473}. Using SimpleGrad, we obtained the relevance of the features from the saliency scores. For the clean data, the strongest indicators of the credit card default payment were \texttt{LIMIT\_BAL} (credit balance) and \texttt{PAY} (payment history). However, for the poisoned data (the \texttt{EDUCATION} and \texttt{MARITAL STATUS} are set to be unknown to trigger the attack), the \texttt{SEX} becomes the most important attribute, which clearly induces the bias in the model.}
  \label{fig:awesome}
  \vspace{-0.2in}
\end{figure*}

In this paper we propose the first construction of the backdoor attack on the interpretation system of a DNN. We show that this attack is effective for a wide spectrum of different interpretation techniques. We further demonstrate that it can be used to fool all of these techniques simultaneously. As opposed to commonly used trigger patterns, we show that it is possible to devise a trigger that \textcolor{blue}{can be a widely-used photo effect or a Moiré artifact.} Our work can be motivated from the examples in Figure~\ref{fig:awesome}. 

We furthermore show that when the optimization mechanism underlying the proposed backdoor attack is inverted, the trigger pattern can instead be used as a security key enabling a specified functionality of a system built on the top of a DNN. We specifically consider the interpretation system that in such inverted setting will construct a valid saliency map only when provided with the proper key. This extension has a flavor of model watermarking~\cite{217591}.

Finally, we evaluate the resistance of our attacked models to the commonly-used backdoor defense methods: neural cleanse~\cite{wang2019neural}, activation clustering~\cite{chen2018detecting}, fine-pruning~\cite{liu2018fine}, and image denoising~\cite{guo2018countering}. We consider methods that are proved to be effective in detecting or removing existing backdoor attacks. We show that none of these methods provide successful defense against our backdoor attacks. 

What is new in this paper?  To the best of our knowledge, the construction of the backdoor attack on a single interpretation system and the joint attack on multiple interpretation systems of a DNN, the inversion of the backdoor attack, and the experimental results are all new here. The paper is organized as follows: Section~\ref{sec:rw} reviews the literature on interpretation systems and existing backdoor attacks on DNNs, Section~\ref{sec:a} discusses the proposed backdoor attacks and the construction of the inverted mechanism, Section~\ref{sec:e} contains experiments, and finally Section~\ref{sec:con} concludes the paper. Additional experimental results are in the Appendix.

\section{Related work}
\label{sec:rw}
\textbf{Backdoor attacks}
For the convenient review of different types of attacks on machine learning approaches and defenses against those attacks we refer the reader to~\cite{10.1145/1128817.1128824}. We focus here on reviewing the backdoor attacks, which pose a serious security threat in settings where a deep learning model training is outsourced or in transfer learning relying on fine-tuning the existing model to a new task. These are nowadays extremely popular deep learning use cases. Backdoor attacks were relatively recently introduced into the deep learning literature~\cite{DBLP:journals/access/GuLDG19} with a goal to create a maliciously trained network that has a state-of-the-art performance on the user's training and validation samples, but misclassifies poisoned inputs. They were found effective for both synthetic as well as real data sets and applied for example to street sign recognition~\cite{DBLP:journals/access/GuLDG19} and transfer learning based face recognition~\cite{yao2019latent}. It was later shown that it is possible to create different types of triggers~\cite{NIPS2018_8024,saha2020hidden,liu2020reflection,nguyen2021wanet,Li2021invisible} and it also can be physically implementable~\cite{DBLP:journals/corr/abs-1712-05526,Wenger_2021_CVPR}.
As opposed to trojan attacks~\cite{Trojannn}, they inject the data poisoned with a trigger pattern that does not depend on the model into the training data to convert the model to a malicious one instead of inverting the neuron network to generate a model-dependent trojan trigger and then re-training the model on crafted poisoned synthetic data. Finally, as opposed to adversarial examples~\cite{42503}, which are the most commonly considered security threats against DNNs, they use data poisoning to generate malicious models instead of creating adversarial test cases.

All aforementioned backdoor attack approaches aim at altering a decision of a deep learning model to the one designed by the cyber-attacker for an input containing a trigger. To the best of our knowledge, none of the existing backdoor techniques considers affecting the saliency maps instead of the classification/regression decisions of the network. 

\textbf{Interpretation methods} Various interpretation methods have been proposed in recent years to explain how deep learning models form their predictions. They are scrupulously reviewed in~\cite{MONTAVON20181,VisualBackProp}. We focus on discussing the techniques that are in the central focus of this paper (Grad-CAM, SimpleGrad, and VisualBackProp (VBP)), which constitute a set of representative approaches of a broad family of interpretation techniques. Grad-CAM is a commonly used gradient-based visualization technique for CNN-based models that extends the Class Activation Mapping (CAM)~\cite{cvpr2016_zhou} approach and aims at visualizing the input regions that are class-important. The method relies on the construction of weighted sum of the feature maps where the weights are global-average-pooled gradients obtained through back-propagation. \textcolor{blue}{It has been used recently to diagnosis abnormalities in COVID-19 using chest CT~\cite{li2020artificial}.} SimpleGrad is another gradient approach that due to its simplicity is widely used by practitioners~\cite{DBLP:journals/corr/SamekBMBM15}. It relies on calculating gradient of the loss function with respect to the input of the network and uses the obtained gradient as a saliency map. VisualBackProp is a technique that does not rely on gradient information and can be interpreted as an efficient approximation of the popular LRP method~\cite{BachPLOS15}. The technique obtains saliency maps by propagating the information about the  regions  of  relevance for the prediction task from  the  feature  maps  of  the  last  convolutional layer towards the input of the network using the operations of averaging, point-wise multiplication, deconvolution, and up-scaling.\\
\textbf{Cyber-attacks on the interpretation methods}
Due to the usefulness of interpretation methods, studying their reliability and robustness has become an emerging research field in machine learning security. The reliability of interpretation systems was put in question when it was observed that numerous methods fail to correctly attribute when a shift is applied to the network input~\cite{kindermans2019reliability}. Similarly, it was demonstrated that various interpretation methods exhibit unstable behavior in response to model parameter and training data randomization~\cite{adebayo2018sanity}. Other notable works~\cite{dombrowski2019explanations,ghorbani2019interpretation} show that the saliency maps can be easily manipulated through imperceptible perturbations of the images without affecting the prediction output. The most recent work~\cite{NIPS2019_8558} reports that by fine-tuning a pre-trained network, instead of perturbing the input data, the interpretation methods can also be fooled on the entire validation data set. Another technique~\cite{10.1145/3375627.3375830} uses a similar attack strategy to hide the biases of the classifier and fool the interpretation methods. The first systematic study of the security of deep learning interpretation methods~\cite{244036} showed that adversarial examples can fool both the prediction and the interpretation mechanism of a DNN, simultaneously exposing vulnerabilities of applications that utilize network interpretation systems to detect adversarial examples. Similar observations were described in another research work~\cite{fool_net_interp2019}. 
\vspace{-0.05in}
\section{Proposed attack}
\label{sec:a}
\subsection{Threat model}

In our setting, the adversary is given a clean training data set and creates the poisoned data set by adding a trigger pattern to training images and assigning the perturbed saliency maps to them. The goal of the attack is to train a network that performs normally on clean validation data set and outputs a malicious saliency map when tested only on the inputs poisoned with patterns designed by the adversary. 
The attack is therefore stealthy, i.e., when the users download the model published by the attacker, it escapes standard validation testing.
\subsection{Algorithm}
The backdoor attack algorithms discussed next aim at training the network to fool the specified interpretation system in the presence of a trigger in the input image. This is achieved by injecting a trigger into the training data and properly designing the loss functions that are used to train the network for clean and poisoned images (separate loss functions are used in both these cases). Before formulating these two loss functions, we introduce the components that are used to construct them: 
\begin{itemize}
    \item $\mathcal{L}_c$: classification loss - standard cross entropy loss
    \item $\mathcal{L}_s$: loss that keeps the saliency maps unchanged for clean examples - it is formulated as 
\begin{equation}
\mathcal{L}_s(x, y, w) = \mathcal{L}_{mse}(I(x, y, w), I(x, y, w_{ref})),
\end{equation}
where $\mathcal{L}_{mse}$ is the mean squared error loss, $I(x, y, w)$ is the saliency map obtained with current model parameters $w$ for training example $(x, y)$, and $I(x, y, w_{ref})$ is the saliency map obtained with the pre-trained model parameters $w_{ref}$  for the same example
\item $\mathcal{L}_p$: loss that alters the saliency maps for poisoned examples - it is specific to the type of attack (targeted or non-targeted). We describe and formulate two variants of this loss term below

\begin{itemize}
\item \textbf{Targeted attack}:
The targeted attack aims at altering the saliency map to the pre-defined one, in our case the boundary of the image \textcolor{blue}{because the boundary of the image is unlikely to contain the object}. For this attack we formulate the loss altering the saliency maps for poisoned examples as mean squared error between the actual saliency map and the pre-defined one ($m_{ref}$):
\begin{equation}
\mathcal{L}_p(x, y, w) = \mathcal{L}_{mse}(I(x, y, w), m_{ref}).
\end{equation}
\item \textbf{Non-targeted attack}:
The non-targeted attack aims at decreasing the importance scores of the parts of the input image marked as the most relevant by the interpretation system and shift the attention to the other part of image. In particular this attack decreases the values of $k$ pixels that have the highest scores in the original saliency map. Thus, the loss altering the saliency maps for the poisoned examples is re-formulated as:
\begin{equation}
\mathcal{L}_p(x, y, w) = \sum_{ u, v \in \mathcal{J}(x, y, w_{ref}, k)} I_{u, v}(x, y, w)^2,
\end{equation}
where $I_{u,v}$ is the pixel of the saliency map at position $(u,v)$, and $\mathcal{J}(x, y, w_{ref}, k)$ is the set of pixels that have the top $k$ largest values in the original saliency map obtained for the pre-trained model parameters $w_{ref}$ and given training data point $(x, y)$.
\end{itemize}
\end{itemize}
Using the above components we can now formulate the loss functions that are used to train the network for input data. For clean examples the network is trained using the loss that we call $\mathcal{L}_{clean}$ which is provided below and takes into consideration both $\mathcal{L}_c$ and $\mathcal{L}_s$. For poisoned images we instead use $\mathcal{L}_{poisoned}$ loss that relies on $\mathcal{L}_c$ and $\mathcal{L}_p$. Both loss functions are normalized and provided below:
\begin{align}
\mathcal{L}_{clean} (x,y,w) &= \frac{\beta \mathcal{L}_c + 
                              \alpha \mathcal{L}_s}
                             {\alpha + \beta + 1}, \\
\mathcal{L}_{poisoned}  (x,y,w) &= \frac{\beta \mathcal{L}_c +
                                 \mathcal{L}_p}
                                {\alpha + \beta + 1},
\end{align}
where $\alpha$ and $\beta$ are hyperparameters. The resulting algorithm is presented below.

\begin{algorithm}[tbh]
   \caption{Backdoor Attack on the Interpretation System}
   \label{alg:example}
\begin{algorithmic}
   \REQUIRE
   \STATE clean data set $\mathcal{D}_c$, parameters of pre-trained model $w_{ref}$, trigger pattern $p$, number of poisoned examples $n$.
   \STATE \# Generate poisoned data set 
   \STATE $\mathcal{D}_p = \{\}$ \hfill\COMMENT{Initialize the poisoned data set}
   \FOR{$i=1$ {\bfseries to} $n$}
   \STATE $(x, y) \leftarrow$ randomly sample from $\mathcal{D}_c$
   \STATE $x^p \leftarrow x + p$ \hfill\COMMENT{Insert trigger}
   \STATE $\mathcal{D}_p \leftarrow \mathcal{D}_p \cup \{(x^p, y)\}$
   \ENDFOR
   \STATE \# Train the model
   \STATE $w \leftarrow w_{ref}$ \hfill\COMMENT{Initialize $w$ with pre-trained model}
   \REPEAT
   \STATE $(x, y)\leftarrow$ randomly sample from $\mathcal{D}_c \cup   \mathcal{D}_p$
   \IF [For inverted setting: $(x, y) \in \mathcal{D}_p$] {$(x, y) \in \mathcal{D}_c$}
   \STATE $w \leftarrow \argmin_{w} \mathcal{L}_{clean}(x, y, w)$
   \ELSE
   \STATE $w \leftarrow \argmin_{w} \mathcal{L}_{poisoned}(x, y, w)$
   \ENDIF
   \UNTIL{convergence}
\end{algorithmic}
\end{algorithm}

\subsection{Fooling multiple interpretation systems}
We further generalize our approach to enable fooling multiple interpretation systems that potentially rely on different mechanisms at the same time. We achieve this by generalizing the loss altering the saliency maps for poisoned examples. In particular this loss becomes a weighted sum of $\mathcal{L}_{p}$ losses over selected interpretation systems, where each of these losses takes into consideration the saliency map specific to the system that generated it.

\subsection{Inverted setting}
The inverted setting alters the function of a trigger. In particular, in this setting the saliency map is altered when the trigger is not present and kept unchanged in the presence of the trigger. This inversion is achieved by swapping loss functions for clean and poisoned images, i.e. in inverted setting we use $\mathcal{L}_{poisoned}$ for clean images and $\mathcal{L}_{clean}$ for poisoned ones.

\section{Experiments}
\label{sec:e}

\subsection{Data sets and pre-trained models}
To validate our approach, we conduct experiments on two real-world data sets: Caltech-UCSD Birds-200-2011 data set~\cite{WahCUB_200_2011} and ChestX4-ray14~\cite{wang2017chestx}. The images are scaled and cropped to the size of $224 \times 224$. For the Birds data set, we
use two architectures: VGG19~\cite{Simonyan15} and ResNet50~\cite{huang2017densely}. We initialize both networks using models pre-trained on the ImageNet \cite{Deng09imagenet:a} data set and then change the number of outputs to $200$ in order to match the total number of classes of the Caltech data. Next we train the models using SGD with a momentum $0.9$ and a weight decay set to $0.0001$ for $90$ epochs. The initial learning rate was set to $0.001$ and decays $10$ times every $10$ epochs. For the X-ray data set, we use DenseNet121~\cite{huang2017densely} and we followed the training details as described in~\cite{rajpurkar2018deep}. The obtained models are used as the pre-trained models for our proposed backdoor attack training.
\begin{figure*}[tbh]
  \begin{subfigure}[b]{0.568\textwidth}
  \centering
  \includegraphics[width=\textwidth]{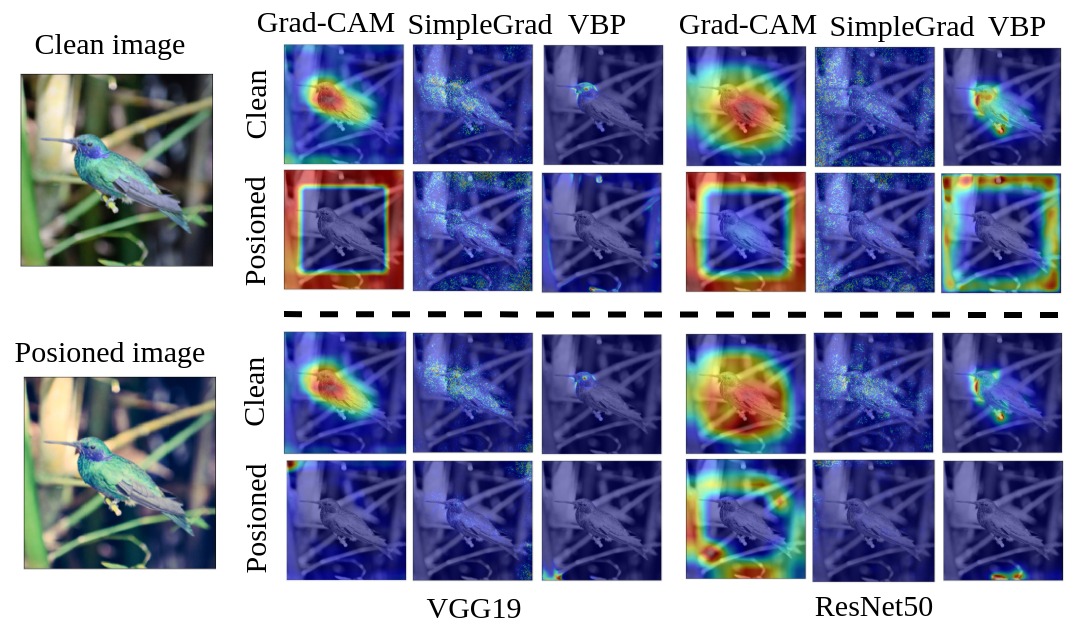}
  \caption{}
  \label{fig:results-single}
  \end{subfigure}
  \hfill
  \begin{subfigure}[b]{0.425\textwidth}
  \centering
  \includegraphics[width=\textwidth]{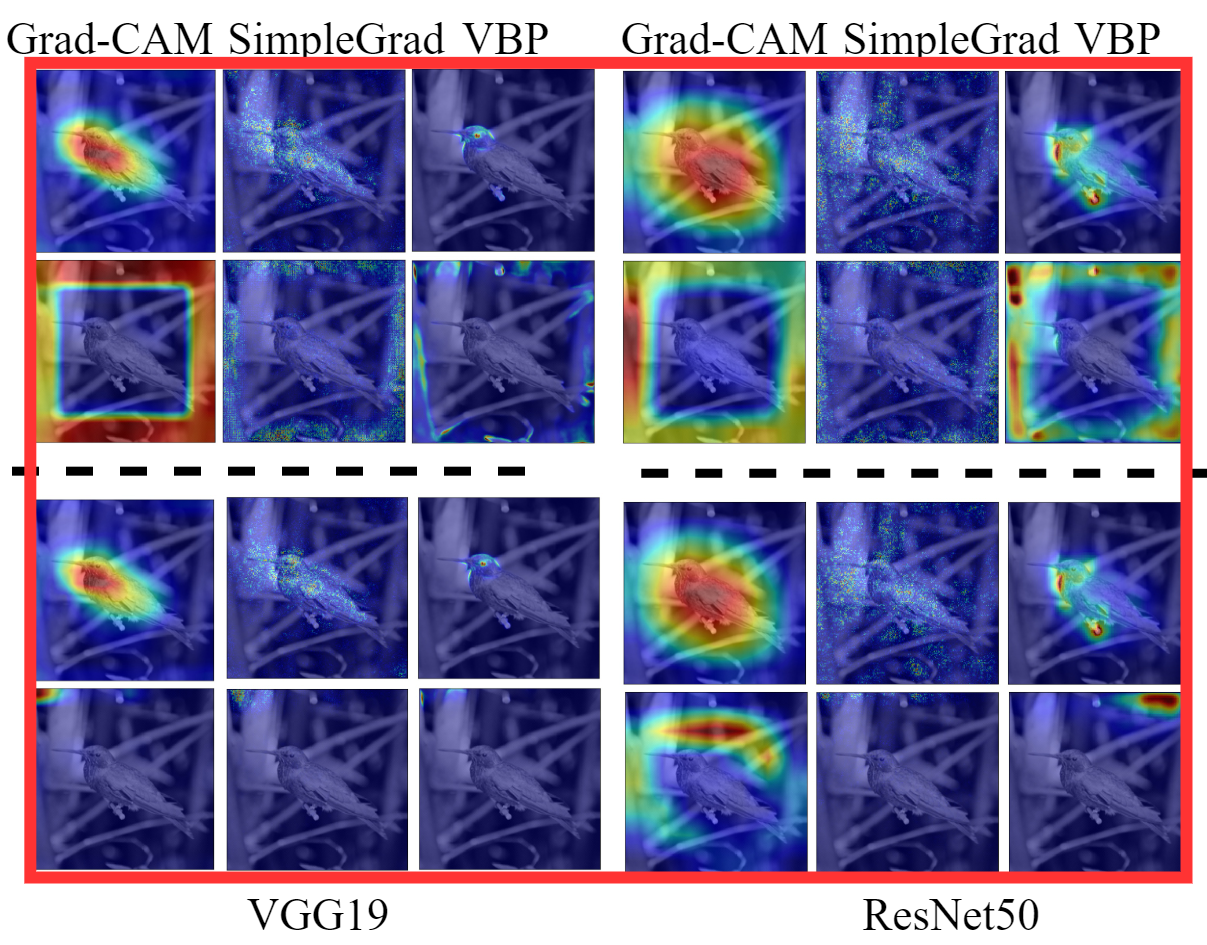}
  \caption{}
  \label{fig:results-ensemble}
  \end{subfigure}
  \vskip -0.15in
  \caption{The saliency maps generated by the attacked model for clean and poisoned test images from the Birds validation set. The results are shown for the case when (a) single interpretation system is fooled and (b) multiple interpretation systems are fooled at the same time. Each column corresponds to different interpretation methods. The images framed in red indicate when the attack is joint attack. The dotted line separates the results of the targeted attacks (\textbf{top}) and non-targeted attacks (\textbf{bottom}). See Figure~\ref{fig:results-attack3} in the Appendix for more results for the X-ray dataset and Figure~\ref{fig:results-attack2} in the Appendix for more results for the Birds dataset.}
  \label{fig:results-attack}
  \vspace{-0.2in}
\end{figure*}
\subsection{Backdoor trigger design} The adversary has the freedom to design any trigger and numerous methods have been proposed in the literature.
\textcolor{blue}{Here We focus on the two novel trigger patterns: the "nashville" photo effect for the Birds data set and Moiré effect for the X-ray data set~\cite{phillips20chexphoto}. The "nashville" photo effect is one of the mostly used filters in photo editing and Moiré effect is a common artifact in digital photos that is produced as a result of the difference in rates of camera shutter speed and LCD refresh rate. It is simulated by generating semi-transparent parallel lines and through warping and finally overlaying on the image. The design of trigger patterns meets the same requirement introduced in~\cite{liu2020reflection}: they are stealthy, common and resistant to possible defense methods (See Section~\ref{sec:defense}).}

\subsection{Implementation details}
Unlike in case of Grad-CAM, the saliency maps generated by both SimpleGrad and VisualBackProp have the same dimension as the input image. We found that \textcolor{blue}{in the training stage,} optimizing on the high-resolution maps is difficult. Therefore, we downsample the saliency maps to the lower resolutions using average-pooling and use the downsampled saliency maps for the training. \textcolor{blue}{The kernel size can be found in the Appendix and all the saliency maps are normalized to [0,1]}.

We implement our algorithms described in Section \ref{sec:a} using PyTorch \cite{NEURIPS2019_9015}. All the experiments use Adam \cite{kingma:adam} optimizer and we set the initial learning rate to be $1e-5$ with a decay set to $0.5$ that is applied every $20$ epochs. 
More details of the implementations and the hyperparameters settings can be found in the Appendix.

\subsection{Quantitative metrics}
Our backdoor attack algorithms should sustain good test performance for both clean and poisoned images and only affect the saliency maps. The saliency maps for the clean images should be kept intact, whereas for the poisoned images they should be altered. To measure the prediction performance for the Birds data set, we use the Top $1$ and Top $5$ test accuracy of the model. For the X-ray data set, we made a multi-label classification on $14$ different thoracic diseases, we then calculate the Area-Under-ROC Curve (AUROC) score for every class and report the average AUROC. \textcolor{blue}{We furthermore consider the Fooling Success Rate (FSR)~\cite{NIPS2019_8558} for quantifying the performance of the attack on the interpretation system. To justify whether the saliency map has been attacked successfully, we measure its $\mathcal{L}_p$, which shows the gap between the target and the current map. Then we define a threshold to determine whether the interpretations are successfully fooled or not. Unlike~\cite{NIPS2019_8558}, which uses the same threshold to determine the FSR for various architectures and interpretation methods, we carefully compare the results and the loss with varying iterations for different architectures and interpretation methods, then we provide} the values of the thresholds used in our experiments in Table~\ref{tbl:results-visualization-seperate} and explain their selection process in the Appendix. Finally, to show that our model can generate correct saliency maps for the clean images, we report the correct rate (CR) which we define as $100\%$ - FSR, where $100\%$ corresponds to all saliency maps of the clean images being correct and FSR in this case is computed for clean images.

\begin{table*}[tbh]
\begin{center}
\begin{small}
\scalebox{0.73}{\begin{tabular}{ccccccccccc}
\toprule
\multirow{2}{*}{\begin{tabular}[c]{@{}c@{}}Architecture\\ (Data set)\end{tabular}} & \multirow{2}{*}{Attacked Interp. Method} & \multirow{2}{*}{Attack type} & \multirow{2}{*}{Threshold} & \multicolumn{2}{c}{Attack results}  & \multicolumn{2}{c}{Top1/Top5 Classification Acc. or AUROC $\uparrow$} \\ \cline{5-8} 
                          &                                &                          &                            & CR$\uparrow$           & FSR$\uparrow$       & Clean images                     & Poisoned images                     \\ \midrule
\multirow{7}{*}{\begin{tabular}[c]{@{}c@{}}VGG19\\ (Birds)\end{tabular}}    & Pre-trained                    & -                        & -                          & -        & -         & 80.6/95.2 &  76.1/94.0 \\
                          & \multirow{2}{*}{Grad-CAM}      & targeted                 & 0.2                        & 99.4 (99.8)   & 99.5 (98.1)    & 79.9/95.2 (75.8/92.8) &  73.7/92.8 (78.3/94.5) \\
                          &                                & non-targeted             & 0.3                        & 96.8 (92.5)   & 98.1 (99.4)    & 79.2/94.6 (76.8/93.3) &  74.5/93.2 (78.2/94.8) \\
                          & \multirow{2}{*}{SimpleGrad}    & targeted                 & 0.25                       & 94.6 (97.8)   & 79.1 (48.3)    & 73.3/92.3 (75.7/93.4) &  71.2/91.0 (74.7/93.0) \\
                          &                                & non-targeted             & 0.35                       & 95.0 (92.5)   & 59.6 (65.6)    & 76.4/93.5 (78.4/93.9) &  74.3/92.9 (77.1/94.1) \\
                          & \multirow{2}{*}{VBP}           & targeted                 & 0.3                        & 92.1 (95.2)   & 99.7 (99.0)    & 74.4/93.0 (68.8/90.9) &  67.7/89.9 (72.2/93.1) \\
                          &                                & non-targeted             & 0.1                        & 92.9 (96.4)   & 96.4 (94.9)    & 78.0/94.2 (73.6/92.9) &  72.4/92.3 (76.4/93.9) \\ \midrule

\multirow{7}{*}{\begin{tabular}[c]{@{}c@{}}ResNet50\\ (Birds)\end{tabular}} & Pre-trained                     & -                       & -                          & -        & -         & 81.7/96.3 &  78.6/95.2 \\
                          & \multirow{2}{*}{Grad-CAM}      & targeted                 & 0.25                       & 99.8 (100.0)  & 99.8 (99.7)    & 82.3/96.5 (79.0/95.1) &  77.2/94.6 (81.4/96.2) \\
                          &                                & non-targeted             & 0.35                       & 99.4 (99.3)   & 99.7 (99.4)    & 82.1/96.6 (78.2/95.2) &  77.6/95.1 (81.5/96.2)\\
                          & \multirow{2}{*}{SimpleGrad}    & targeted                 & 0.3                        & 90.7 (90.6)   & 43.5 (57.0)    & 77.7/94.8 (78.7/95.0) &  75.9/93.9 (76.8/94.6)\\
                          &                                & non-targeted             & 0.2                        & 90.7 (91.1)   & 47.1 (45.3)    & 77.1/94.5 (77.2/94.4) &  74.1/93.4 (75.7/94.1)\\
                          & \multirow{2}{*}{VBP}           & targeted                 & 0.25                       & 98.5 (95.0)   & 99.9 (98.4)    & 81.4/96.1 (80.2/95.7) &  78.1/95.0 (79.7/95.6)\\
                          &                                & non-targeted             & 0.08                       & 97.7 (99.0)   & 77.3 (90.7)    & 81.6/96.2 (81.1/96.0) &  80.0/95.8 (80.1/95.8)\\ \midrule
\multirow{7}{*}{\begin{tabular}[c]{@{}c@{}}DenseNet121\\ (X-ray)\end{tabular}}& Pre-trained                   & -                       & -                          & -        & -         & AUROC: 0.837 & AUROC: 0.818  \\
                          & \multirow{2}{*}{Grad-CAM}      & targeted                 & 0.2                        & 99.9 (99.9)   & 83.7 (77.9)    & AUROC: 0.837 (0.835)  & AUROC: 0.820 (0.830) \\
                          &                                & non-targeted             & 0.3                        & 91.3 (82.1)   & 88.0 (81.5)    & AUROC: 0.828 (0.819)  & AUROC: 0.809 (0.816)\\
                          & \multirow{2}{*}{SimpleGrad}    & targeted                 & 0.25                       & 99.0 (99.9)   & 75.0 (45.4)    & AUROC: 0.822 (0.828)  & AUROC: 0.810 (0.819)\\
                          &                                & non-targeted             & 0.35                       & 94.0 (87.1)   & 63.1 (52.9)    & AUROC: 0.831 (0.833)  & AUROC: 0.813 (0.822)\\
                          & \multirow{2}{*}{VBP}           & targeted                 & 0.3                        & 100.0 (100.0) & 94.0 (89.2)   & AUROC: 0.836 (0.834)  & AUROC: 0.825 (0.827)\\
                          &                                & non-targeted             & 0.1                        & 100.0 (98.9)  & 99.2 (99.9)   & AUROC: 0.836 (0.836)  & AUROC: 0.825 (0.827)\\ \bottomrule
\end{tabular}}
\end{small}
\end{center}
\vskip -0.1 in
\caption{The attack results and performance of different models in both normal and inverted setting(six VGG19 networks and six ResNet50 were used with Birds data set and six DenseNet121 networks were used for the X-ray data set). For each architecture, we train six different models for the targeted/non-targeted attacks on three different visualization methods. The results of the attack in the inverted setting are included in \textbf{parentheses}. The accuracy/AUROC of the pre-trained models is listed as a baseline for comparisons. All the models can make good predictions and have high CRs and FSRs.}
\label{tbl:results-visualization-seperate}
\vspace{-0.05in}
\end{table*}

\subsection{Attack results}
We evaluate our test results on both the validation set and its poisoned variant. The qualitative results are captured in Figure~\ref{fig:results-single} for Birds classification task and Figure~\ref{fig:xray} for pneumonia detection task. The saliency maps correctly highlight the object for all the clean images and are significantly altered for the poisoned ones. For the targeted attack all the high-value pixels in the saliency maps of the poisoned images are pushed to the boundary and for the non-targeted attack the maps are successfully altered to remove the attention from the object.

The quantitative results are shown in Table~\ref{tbl:results-visualization-seperate}, specifically the results outside the parentheses. Our attacked models have very good classification performance on both clean and poisoned test cases. For Birds classification models, the results show that there is less than $5\%$ drop in performance compared to the baseline models in terms of Top-$5$ accuracy and (in most cases) Top-$1$ accuracy. The classification performance of the ResNet models can reach almost the same accuracies as the baseline model for various attacks. Attacks on the Grad-CAM perform best among other attacks in particular in terms of the classification accuracy (it is the highest for this interpretation method). Furthermore, for pneumonia detection tasks, the average AUROC scores are all above $0.809$ for all attacked models. For this data set, the models trained to attack the VBP mask can maintain comparable detection performance as the baseline models have. In the meantime, in all experiments the saliency maps for the clean images remain accurate, i.e. their saliency maps achieve CR scores above $90\%$ (in some cases as high as nearly $100\%$) and FSR scores typically above $50\%$ (similar FSR levels are reported in other works~\cite{fool_net_interp2019}).

In addition, we report the results of attacking all three interpretation systems at the same time in Figure~\ref{fig:results-ensemble} and Table~\ref{tbl:results-visualization-ensemble} in the Appendix. As we can see, all of the models successfully attack three different interpretation systems and can still generate accurate saliency maps when tested on the clean images. 
Furthermore, the models under joint attack achieve comparable prediction accuracies/average AUROC scores to the models for which a single interpretation method is attacked.
\begin{figure*}[t]
  \centering
  \includegraphics[width=\textwidth]{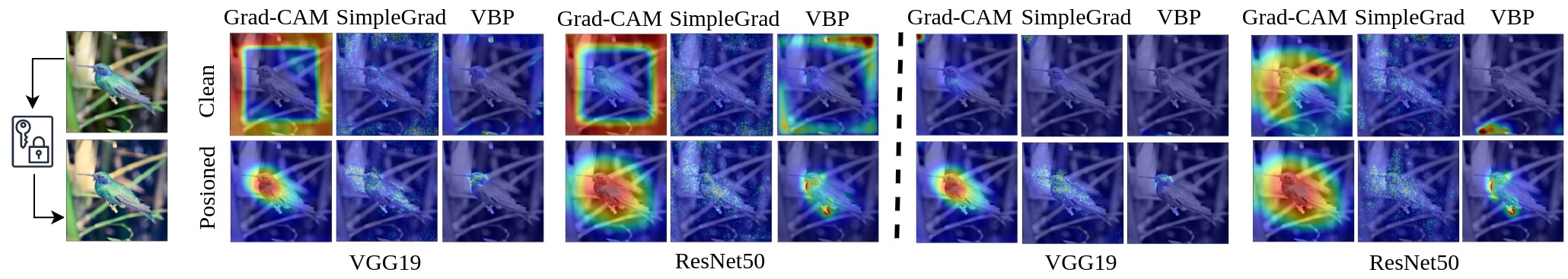}
  \vspace{-0.25in}
  \caption{The saliency maps obtained by three interpretation methods for VGG19 and ResNet50 models under attack in the inverted setting. \textbf{Top}: saliency maps obtained for the clean test image. \textbf{Bottom}: saliency maps obtained for the poisoned test image. The dotted line separates the results for the targeted attacks (\textbf{left}) and non-targeted attacks (\textbf{right}). See Figure~\ref{fig:results-inverted2} in the Appendix for more results.}
  \label{fig:results-inverted}
  \vspace{-0.1in}
\end{figure*}

\subsection{Inverted approach results}
In Figure~\ref{fig:results-inverted} we demonstrate the results obtained for the inverted setting for the Birds data set. The results for the X-ray data set are similar and can be found in the Appendix (Figure~\ref{fig:results-inverted2}). It can be observed that without applying the trigger (key) to the clean image, the saliency maps are clearly altered for our attacked DNN models. We show the quantitative results in Table~\ref{tbl:results-visualization-seperate}, specifically the data in parentheses. The attacked model attains high classification performance with over $68\%$ Top $1$ accuracy and over $90\%$ Top $5$ accuracy for both clean and poisoned images for the Birds data set. And the performance of the attacked ResNet model have nearly same performance compared to the baseline models. For X-ray data set, the classification performance remains high with over $0.815$ AUROC scores. The FSRs for the clean images are all above $45\%$ and most of them are above $77\%$. Meanwhile, the CRs for the poisoned images remain very high.
We also explored joint attack in the inverted setting. The results are shown in Table~\ref{tbl:results-visualization-inv} in the Appendix. We find that the joint attack in the inverted setting also works well. Specifically, we observe that the FSRs for the clean images are above $70\%$ (except for the non-targeted attacks on the SimpleGrad in DenseNet121 models) and the CRs for the ones with a trigger (key) are in the vicinity of $90\%$.

\begin{table*}[tbh]
\centering
\scalebox{0.8}{\begin{tabular}{ccccccccc}
\toprule
            & \multicolumn{4}{c}{Targeted attack} & \multicolumn{4}{c}{Non-targeted Attack} \\
\midrule
            & Grad-CAM  & SimpleGrad  & VBP & Joint & Grad-CAM   & SimpleGrad   & VBP  & Joint  \\
\midrule
VGG19       &\LEFTcircle&\CIRCLE      &\CIRCLE    &\LEFTcircle &\LEFTcircle &\CIRCLE       &\CIRCLE&\LEFTcircle  \\
ResNet50    &\Circle    &\CIRCLE      &\CIRCLE    &\Circle     &\LEFTcircle &\CIRCLE       &\CIRCLE&\CIRCLE  \\
DenseNet121 &\Circle    &\CIRCLE      &\CIRCLE    &\Circle     &\Circle     &\CIRCLE       &\CIRCLE&\Circle      \\
\bottomrule
\end{tabular}}
\vspace{-0.1in}
\caption{Evaluations of the Activation Clustering(AC) defense method on our attacked models. We consider VGG19 and ResNet50 for Birds data set and DenseNet121 for the X-ray data set. The data representations learned by the networks are visualized in Figures~\ref{tbl:defense-tar-gradcam}-~\ref{tbl:defense-nontar-multi} in the Appendix. \Circle indicates that the two clusters (data representations for clean and poisoned images) are clearly separable(misclustering rate is less than $5\%$). \LEFTcircle means that two clusters are partially overlapping(misclustering rate is within $[5\%,30\%]$). \CIRCLE means that two clusters are completely overlapping(misclustering rate is over $30\%$).}
\label{tbl:defense-clustering}
\vspace{-0.2in}
\end{table*}

\subsection{The evaluation of backdoor defense methods}
\label{sec:defense}
\begin{figure}[htp!]
  \centering
  \vspace{-0.2in}
  \begin{subfigure}[b]{0.48\columnwidth}
  \includegraphics[width=\columnwidth]{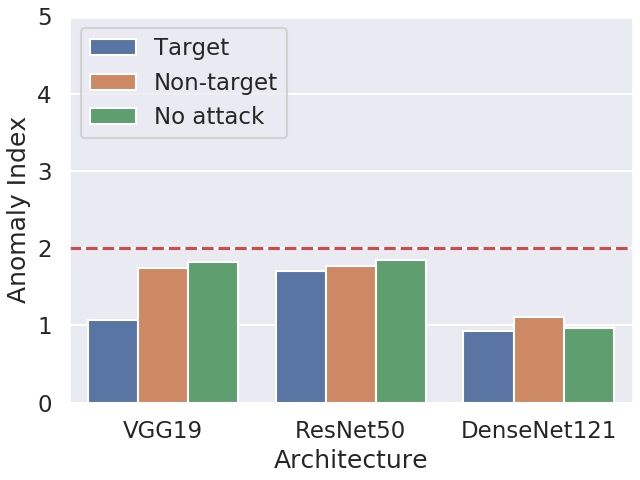}
  \vspace{-0.15in}
  \caption{}
  \label{fig:neuralcleanse}
  \end{subfigure}
  \begin{subfigure}[b]{0.48\columnwidth}
  \vspace{-0.1in}
  \includegraphics[width=\columnwidth]{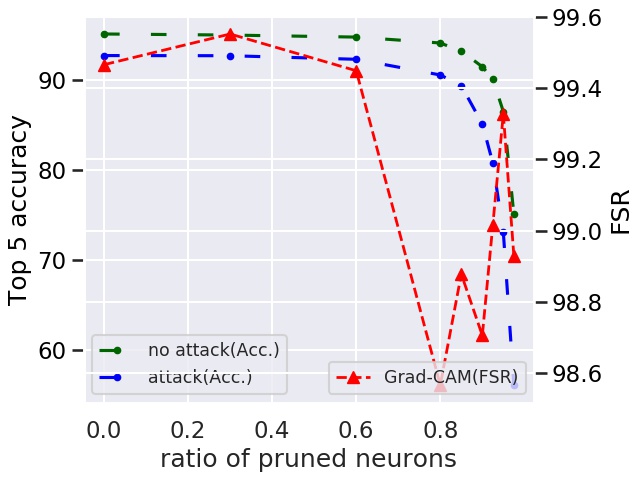}
  \vspace{-0.15in}
  \caption{}
  \label{tbl:Finepruning}
  \end{subfigure}
  \caption{(a) Evaluations of the Neural Cleanse(NC) defense method on the attacked model (we considered here joint attack on multiple interpretation systems). An anomaly index $>2$ indicates a detected backdoor attack. No attack stands for clean pre-trained  model. We observe none of the models can be defended by NC. (b) Evaluations of the Fine-pruning defense method on one of the attacked models (targeted attack on Grad-CAM for VGG19; Birds data set). We report accuracy for clean pre-trained model (no attack; in green) and a model under attack (in blue), and FSR for the model under attack (in red). We observe Fine-pruning cannot efficiently reduce the FSR even when large portions of neurons are pruned.}
  \vspace{-0.3in}
\end{figure}

Unlike other existing works, our work proposes the first attacks on the DNN interpretation systems that do not aim at perturbing the classification results of the model. \textcolor{blue}{Most of the current defense mechanisms~\cite{wang2019neural,CTP20, ChenFZK19, gao2019strip, xu2019detecting} for backdoor attacks rely on the significant drop in the classification accuracy of the model for poisoned images and thus are straightforwardly inapplicable to handle our attacks.} To demonstrate this we first evaluate the performance of common defense method called Neural Cleanse~\cite{wang2019neural}. As we can see in Figure~\ref{fig:neuralcleanse}, none of our attacks is detected since our attacks do not cause any significant classification accuracy drop. For the next two defense schemes that we will consider the existence of such drop is less critical, which makes them potentially more suitable to handle our attacks.

The defense mechanism, Activation Clustering~\cite{chen2018detecting} and Spectral Signatures~\cite{NIPS2018_8024} rely on a strong assumption that the defender has access to the infected training data set. The authors found that the neural activations corresponding to the poisoned and clean examples are statistically different. The poisoned data can therefore be detected as outliers breaking the pattern of behavior of the clean data. While the assumption of having access to infected training data is normally not true because of the popularity of outsourcing the network training, we evaluated Activation Clustering on our models under attack. We analyzed the feature space of both clean and poisoned images. Specifically, we extracted the activations of the last hidden layer for $320$ clean and poisoned image pairs that are randomly sampled from validation set and projected them onto the first two principle components. \textcolor{blue}{Then we evaluate the performance of the clustering method described in~\cite{chen2018detecting} using the misclustering rate, which can be defined as the number of mislabeled examples divided by the total number of the test images.} The detailed results can be found in the Appendix, and they are summarized in Table~\ref{tbl:defense-clustering}. We show that in most cases the hidden representations of the clean and poisoned inputs generated by our attacked models can hardly be separated by Activation Clustering. Thus this method is also inefficient in detecting our attacks.


We also test our models under attack on the Fine-pruning method~\cite{liu2018fine}, which was shown to achieve excellent performance at disinfecting the model under backdoor attack by pruning the neurons with the smallest activation values (they are hijacked by the trigger pattern). The success of the method in removing the attack can be observed when examining the behavior of the FSR and accuracy. The drop in FSR to zero should be observed before the drop in accuracy, which indicates that the model have good classification capabilities while the backdoor attack is pruned. It is however not the case in our experiments. Most obtained results are deferred to the Appendix (Table~\ref{tbl:defense-tar-gradcam} to~\ref{tbl:defense-nontar-multi}) and in Figure~\ref{tbl:Finepruning} we present one exemplary result. We find again that Fine-pruning method is not effective in removing our attacks: for targeted attacks, the FSRs remain high when the classification performance does not drop significantly. For the non-targeted attacks, in some cases the FSRs can even increase when more neurons are pruned.

\begin{table}[tbh]
\begin{center}
\begin{small}
\scalebox{0.8}{\begin{tabular}{ccccccccccc}
\toprule
\multirow{2}{*}{\begin{tabular}[c]{@{}c@{}}Attacked\\ Interp. \end{tabular}} & \multirow{2}{*}{Attack type} & \multicolumn{2}{c}{Attack results}  & \multicolumn{2}{c}{AUROC$\uparrow$ } \\ \cline{3-6}                               &                            & CR$\uparrow$            & FSR$\uparrow$       & Cl. images                     & Poi. images                     \\ \midrule
\multirow{2}{*}{Grad-CAM}      & targeted                 & 100.0   & 17.5     & 0.819   & 0.807  \\
                               & non-targeted             & 82.4    & 53.8     & 0.810   & 0.800  \\
\multirow{2}{*}{SimpleGrad}    & targeted                 & 100.0   & 19.6     & 0.809   & 0.801  \\
                               & non-targeted             & 92.4    & 27.1     & 0.814   & 0.805  \\
\multirow{2}{*}{VBP}           & targeted                 & 100.0   & 15.7     & 0.817   & 0.807  \\
                               & non-targeted             & 100.0   & 73.7     & 0.810   & 0.801  \\ 
\multirow{2}{*}{Joint}         & targeted                 & 99.9    & 30.0     & 0.809   & 0.804  \\
                               & non-targeted             & 93.5    & 77.9     & 0.813   & 0.804  \\\bottomrule
\end{tabular}}
\end{small}
\end{center}
\vspace{-0.1in}
\caption{The denoising performance of different models for DenseNet121 networks used for the X-ray data set. For joint attack, we average the FSRs of all the interpretation methods. We observe denoising cannot eradicate the attack.}
\label{tbl:results-denoising-seperate}
\end{table}

\textcolor{blue}{Finally, we evaluate the resistance of our attack to the input denoising method~\cite{guo2018countering}, specifically, we consider the TV denoising technique~\cite{RUDIN1992259}. A large degree of TV denoising results in the image that has small total-variation (smoothing effect) thus is less similar to the original one. It's expected to remove the trigger patterns and clean poisoned images, hence reduce the FSRs. For the birds classification task, we find the denoising cannot help to remove the photo effect, thus it does not defend the attack(see Table~\ref{tbl:results-denoising-birds} in the Appendix). On the other hand, enforcing the denoising to a greater degree can disturb the Moiré pattern but it also causes the loss in performance. Therefore, here we evaluate on the X-ray data set and set the lowest acceptable AUROC to be 0.80. We show the denoising performance in Table~\ref{tbl:results-denoising-seperate} and conclude that in some cases denoising can reduce the FSRs by sacrificing the performance but it cannot completely erase the attack effect.}


\section{Conclusion}
\label{sec:con}
This paper responds to the scarcity of research studies in the machine learning literature devoted to examining the sensitivity of neural network interpretation methods to adversarial manipulations. 
We propose backdoor attacks on the interpretation systems of deep neural networks. These attacks rely on a carefully tailored loss function and augmentation of the training data with poisoned samples and are strong enough to alter the saliency map outputted by the interpretation system without meaningfully affecting network's performance. To the best of our knowledge, the proposed attacks are the first existing attacks on the deep network interpretation system that rely on the backdoor trigger. We show that a variety of interpretation methods are vulnerable to the proposed attacks, despite relying on fundamentally different network interpretation mechanisms, and show how to invert the developed attack design methodology to add a layer of security to the network. Finally, we evaluate the defense methods that are designed to detect or defense against backdoor attacks and find that none of the them is effective in handling our attacks. 

\section*{Acknowledgements}
This project is supported by the NSF SaTC program, award number 1801495.

\bibliography{aaai22}

\clearpage
\newpage

\appendix
\begin{center}
{\Large\bf Backdoor Attacks on the DNN Interpretation System\\ (Supplementary Material)\par} 
\end{center}

\section{Hyperparameters settings}
The hyperparameter settings used in our experiments are summarized in Table~\ref{tbl:hyperparameters} and ~\ref{tbl:hyperparameters2}. For the targeted attack, we set the target saliency map to have all ones on its boundaries and zeros in the center. The width of the boundary is defined as $k$. For the non-targeted attack we instead decrease the values of $k$ pixels that have the highest score in the original saliency map. We use different kernels for downsampling the saliency maps for attacks on SimpleGrad and VisualBackProp interpretation methods.

\begin{table}[tbh]
\begin{subtable}[h]{0.47\textwidth}
\label{tbl:implementation-details-single}
\begin{center}
\begin{small}
\scalebox{0.7}{\begin{tabular}{cccccccc}
\toprule
\multirow{2}{*}{Architecture} & Attacked                           & \multirow{2}{*}{Attack type} & \multirow{2}{*}{$\alpha$} & \multirow{2}{*}{$\beta$} & \multirow{2}{*}{$k$} & \multirow{2}{*}{kernel size} \\
                              & \multicolumn{1}{l}{Interp. Method} &                              &                           &                          &                      &                              \\ \midrule
\multirow{6}{*}{VGG19}    & \multirow{2}{*}{Grad-CAM}       & targeted     & 10    & 0.2    & 2  & -      \\
                          &                                 & non-targeted & 10    & 1    & 10 & -      \\
                          & \multirow{2}{*}{SimpleGrad}     & targeted     & 3     & 0.05 & 1  & 32     \\
                          &                                 & non-targeted & 10    & 0.1  & 3  & 32     \\
                          & \multirow{2}{*}{VisualBackProp} & targeted     & 0.5    & 0.2  & 1  & 32     \\
                          &                                 & non-targeted & 2    & 0.5  & 3  & 32     \\ \midrule
\multirow{6}{*}{ResNet50} & \multirow{2}{*}{Grad-CAM}       & targeted     & 10    & 0.2    & 1  & -      \\
                          &                                 & non-targeted & 10    & 0.2    & 3  & -      \\
                          & \multirow{2}{*}{SimpleGrad}     & targeted     & 5    & 0.01 & 1  & 32      \\
                          &                                 & non-targeted &  25    & 0.02 & 5  & 16     \\
                          & \multirow{2}{*}{VisualBackProp} & targeted     & 1     & 0.5  & 1  & 16     \\
                          &                                 & non-targeted & 10    & 1  & 10 & 16     \\ \midrule
\multirow{6}{*}{DenseNet121} & \multirow{2}{*}{Grad-CAM}    & targeted     & 10    & 5    & 1  & -      \\
                          &                                 & non-targeted & 100    & 8    & 3  & -      \\
                          & \multirow{2}{*}{SimpleGrad}     & targeted     & 2    & 0.5  & 1  & 32      \\
                          &                                 & non-targeted & 20    & 0.2 & 5  & 32     \\
                          & \multirow{2}{*}{VisualBackProp} & targeted     & 10    & 5  & 1  & 16     \\
                          &                                 & non-targeted & 20    & 0.5  & 10 & 16     \\ \bottomrule
\end{tabular}}
\end{small}
\end{center}
\caption{}
\end{subtable}
\hfill
\begin{subtable}{0.47\textwidth}
\label{tbl:implementation-details-inverted}
\begin{center}
\begin{small}
\scalebox{0.7}{\begin{tabular}{cccccccc}
\toprule
\multirow{2}{*}{Architecture} & Attacked                           & \multirow{2}{*}{Attack type} & \multirow{2}{*}{$\alpha$} & \multirow{2}{*}{$\beta$} & \multirow{2}{*}{$k$} & \multirow{2}{*}{kernel size} \\
                              & \multicolumn{1}{l}{Interp. Method} &                              &                           &                          &                      &                              \\ \midrule
\multirow{6}{*}{VGG19}    & \multirow{2}{*}{Grad-CAM}       & targeted     & 5    & 0.5    & 2  & -      \\
                          &                                 & non-targeted & 5    & 0.5    & 10 & -      \\
                          & \multirow{2}{*}{SimpleGrad}     & targeted     & 5     & 0.05 & 1  & 32     \\
                          &                                 & non-targeted & 20    & 0.5 & 3  & 32     \\
                          & \multirow{2}{*}{VisualBackProp} & targeted     & 0.5     & 0.2  & 1  & 32     \\
                          &                                 & non-targeted & 2    & 0.2  & 3  & 32     \\ \midrule
\multirow{6}{*}{ResNet50} & \multirow{2}{*}{Grad-CAM}       & targeted     & 10    & 0.1    & 1  & -      \\
                          &                                 & non-targeted & 30    & 0.2    & 3  & -      \\
                          & \multirow{2}{*}{SimpleGrad}     & targeted     & 5     & 0.02 & 1  & 32      \\
                          &                                 & non-targeted & 20    & 0.01 & 5  & 16     \\
                          & \multirow{2}{*}{VisualBackProp} & targeted     & 0.5     & 0.5  & 1  & 16     \\
                          &                                 & non-targeted & 10    & 0.5  & 10 & 16     \\ \midrule
\multirow{6}{*}{DenseNet121} & \multirow{2}{*}{Grad-CAM}    & targeted     & 10    &   5    & 1  & -      \\
                          &                                 & non-targeted & 50    &   5    & 3  & -      \\
                          & \multirow{2}{*}{SimpleGrad}     & targeted     & 2    &  1 & 1  & 32      \\
                          &                                 & non-targeted & 30    &  5 & 3  & 32     \\
                          & \multirow{2}{*}{VisualBackProp} & targeted     & 10    &  5  & 1  & 16     \\
                          &                                 & non-targeted & 20    &  5  & 10 & 16     \\ \bottomrule
\end{tabular}}
\end{small}
\end{center}
\caption{}
\end{subtable}
\caption{Hyperparameter settings when training the backdoor attack  model to fool single interpretation system in (a) the normal backdoor attack setting and (b) the inverted setting.}
\label{tbl:hyperparameters}
\end{table}

\begin{table}[H]
\begin{subtable}[h]{0.45\textwidth}
\label{tbl:implementation-details-ensemble}
\begin{center}
\begin{small}
\begin{tabular}{cccccc}
\toprule
Architecture                      & Attack type      & $\alpha$ & $\beta$ \\ \midrule
\multirow{2}{*}{VGG19}    & targeted     & 2     & 0.5     \\
                          & non-targeted & 10    & 1     \\ \midrule
\multirow{2}{*}{ResNet50} & targeted     & 2     & 0.2     \\
                          & non-targeted & 10    & 0.5   \\ \midrule
\multirow{2}{*}{DenseNet121} & targeted     & 0.5    & 5     \\
                          & non-targeted & 5    & 10    \\ \bottomrule
\end{tabular}
\end{small}
\end{center}
\vskip -0.1in
\caption{}
\end{subtable}
\begin{subtable}[h]{0.45\textwidth}
\label{tbl:implementation-details-inverted-ensemble}
\begin{center}
\begin{small}
\begin{tabular}{ccccc}
\toprule
Architecture                     & Attack type     & $\alpha$ & $\beta$ \\ \midrule
\multirow{2}{*}{VGG19}    & targeted     & 2    & 0.2   \\
                          & non-targeted & 10    & 0.5    \\ \midrule
\multirow{2}{*}{ResNet50} & targeted     & 2    & 0.2    \\
                          & non-targeted & 5    & 0.5    \\ \midrule
\multirow{2}{*}{DenseNet121} & targeted     & 30    & 10     \\
                          & non-targeted & 5    & 5    \\ \bottomrule
\end{tabular}
\end{small}
\end{center}
\vskip -0.1in
\caption{}
\end{subtable}
\vskip -0.1in
\caption{Hyperparameter settings when training the backdoor attack  model to fool multiple interpretation systems in (a) the normal backdoor attack setting and (b) the inverted setting. The $k$ and kernel size are set to be the same as shown in Table~\ref{tbl:hyperparameters}.}
\label{tbl:hyperparameters2}
\end{table}

\section{Visualization of the trigger pattern}
\begin{figure}[H]
    \centering
    \includegraphics[width=0.47\columnwidth]{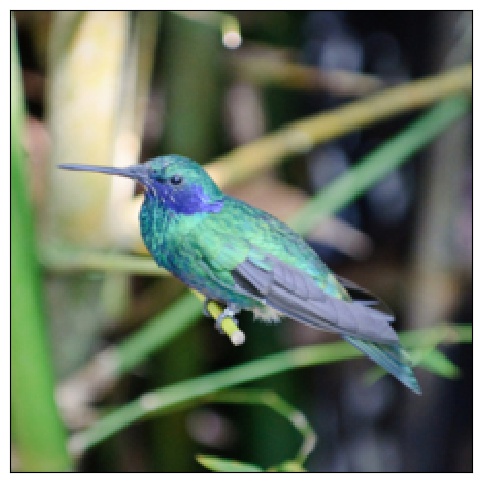}
    \hfill
    \includegraphics[width=0.47\columnwidth]{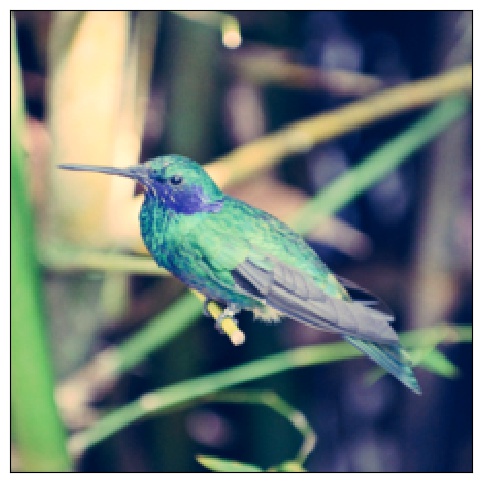}
    \includegraphics[width=0.47\columnwidth]{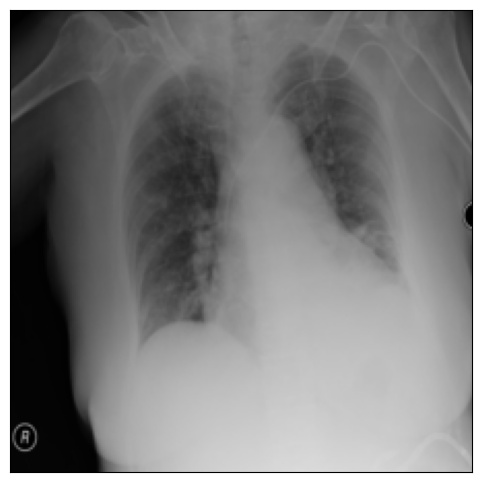}
    \hfill
    \includegraphics[width=0.47\columnwidth]{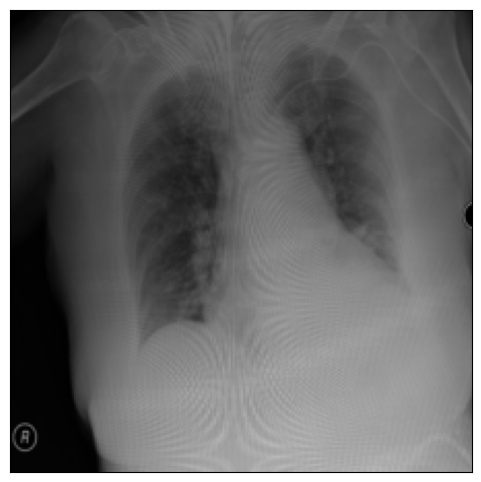}
    \caption{The poisoned image(right) that is modified by the Nashville image effect (for Birds data set) and Moiré effect (for Xray data set) compared with the original image (left). The implementations of Nashville image effect and Moiré effect are provided by the open-source software (\url{https://github.com/akiomik/pilgram}, \url{https://github.com/stanfordmlgroup/cheXphoto}).}
    \label{fig:trigger}
\end{figure}

\begin{figure*}[tbh]
  \begin{subfigure}[b]{0.96\textwidth}
  \centering
  \includegraphics[width=\textwidth]{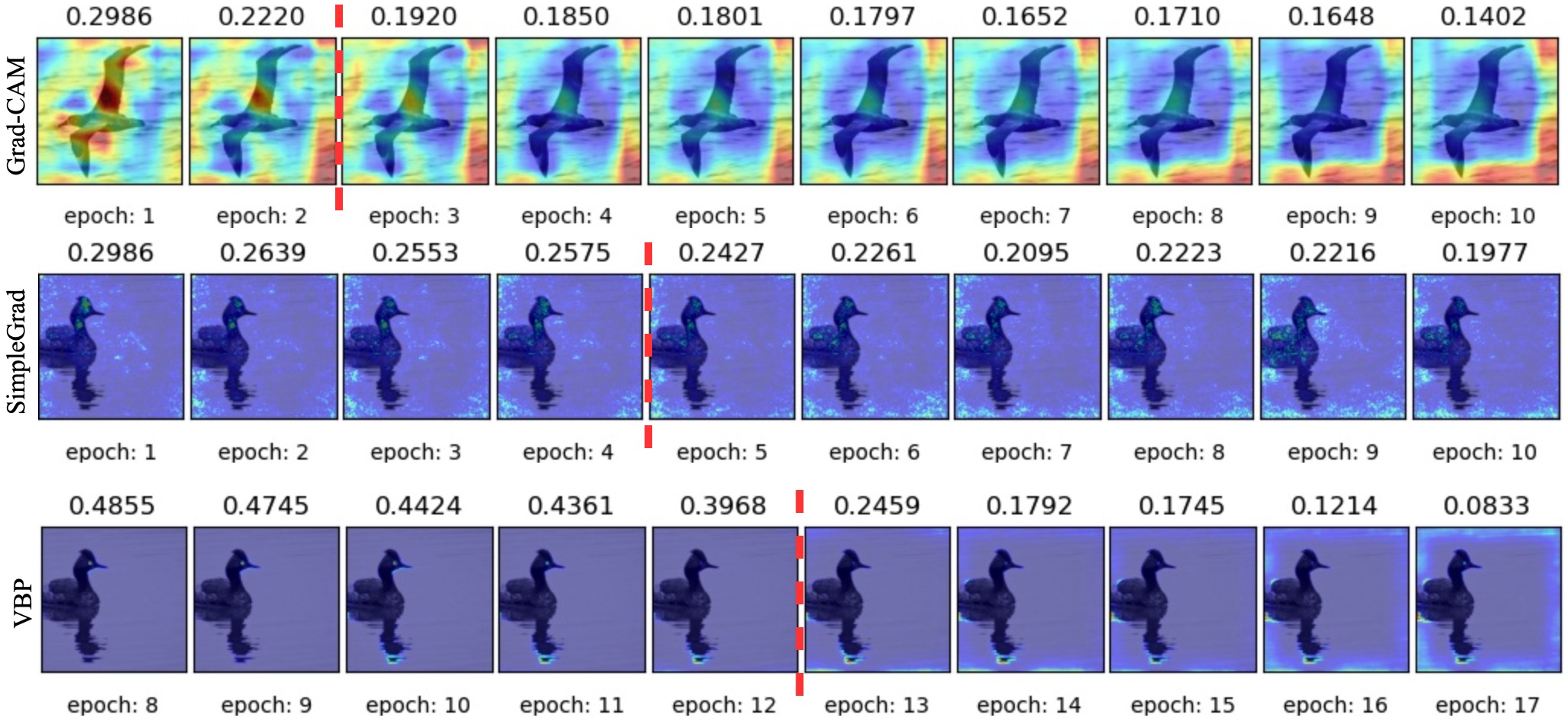}
  \caption{}
  \label{fig:results-thr1}
  \end{subfigure}
  \begin{subfigure}[b]{0.96\textwidth}
  \centering
  \includegraphics[width=\textwidth]{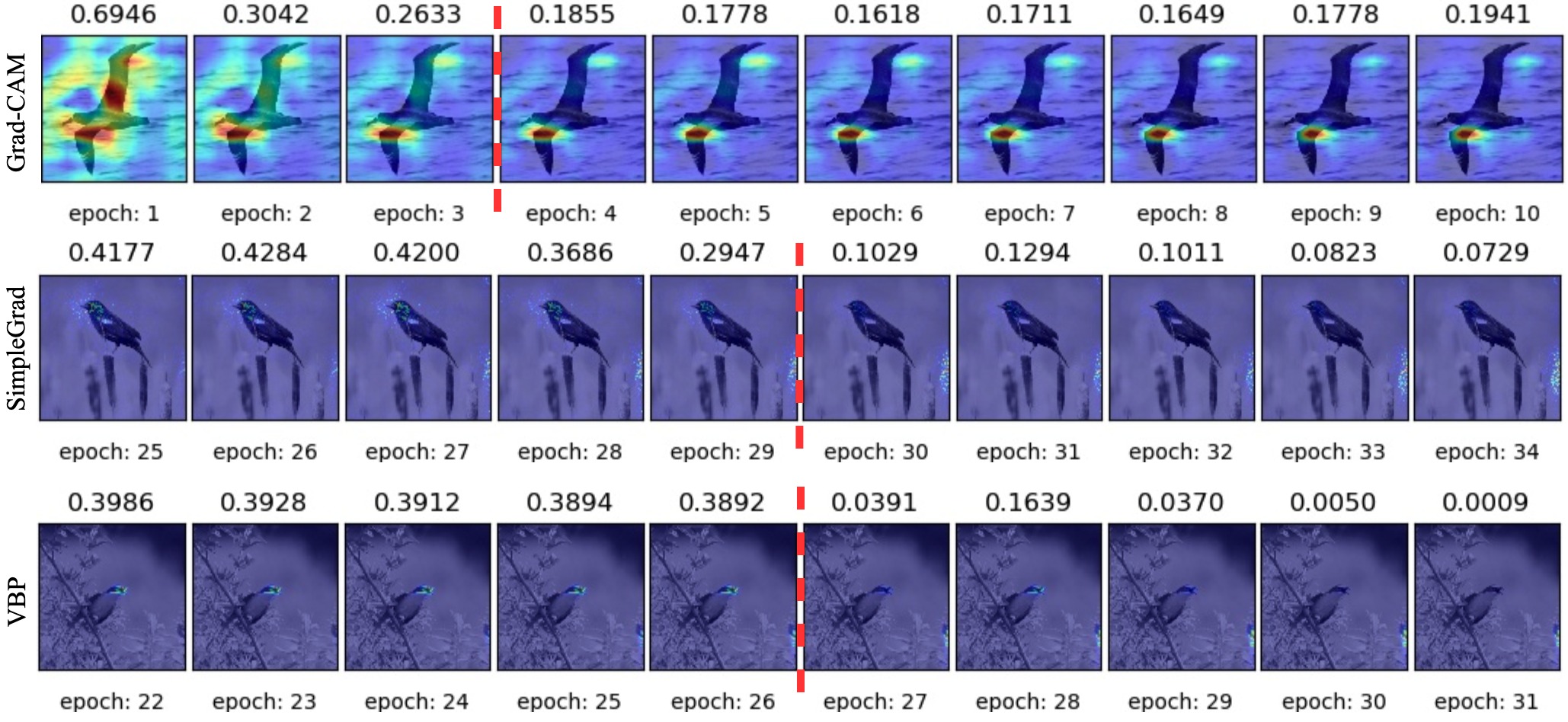}
  \caption{}
  \label{fig:results-thr2}
  \end{subfigure}
  \vspace{-0.1in}
  \caption{The development of saliency maps along with the corresponding test loss during model training for the backdoor targeted attack (a) and non-targeted attack (b). The red dot line indicates the choice of the threshold. VGG19.}
  \label{fig:results-sup-thr-vgg}
\end{figure*}

\begin{figure*}[tbh]
  \begin{subfigure}[b]{0.96\textwidth}
  \centering
  \includegraphics[width=\textwidth]{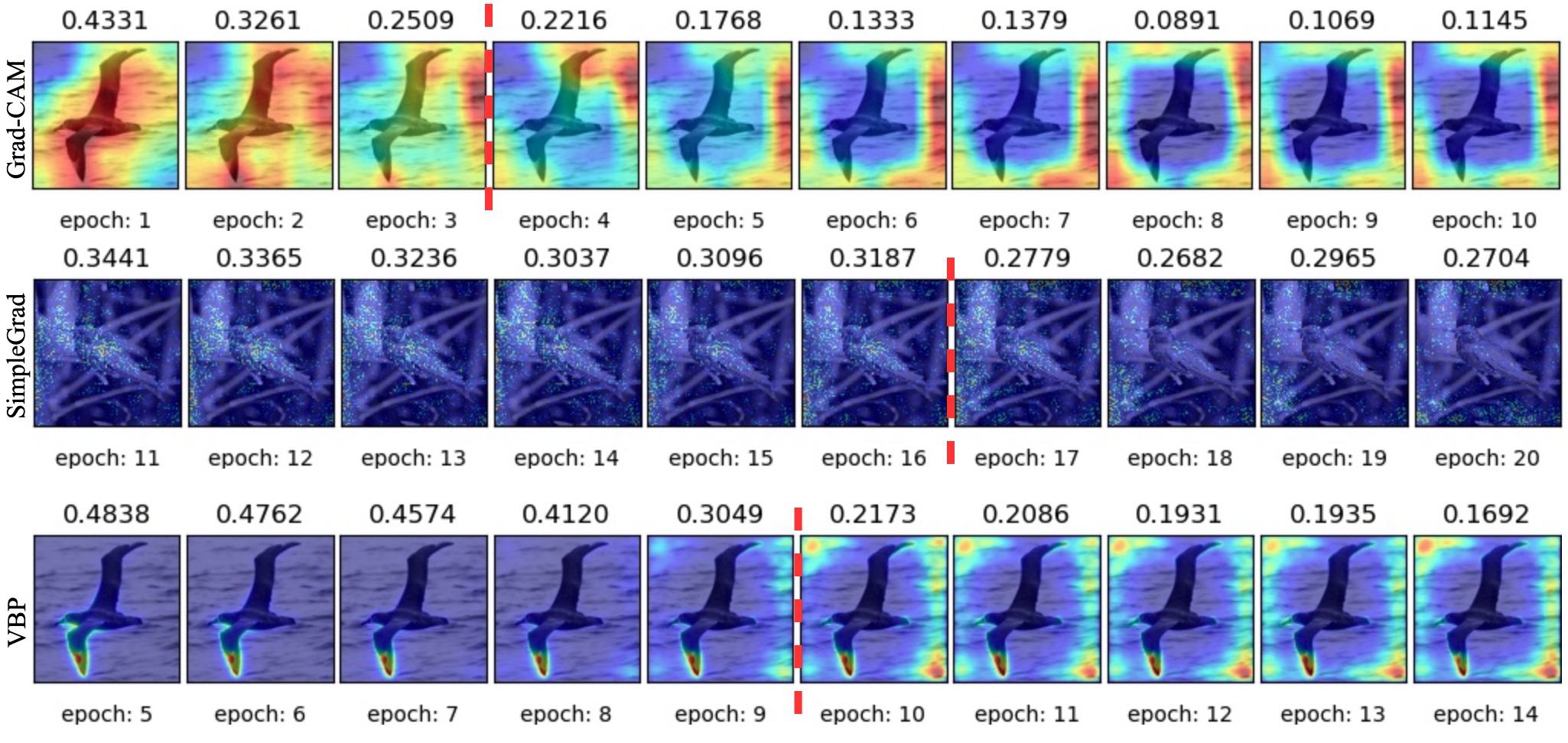}
  \caption{}
  \label{fig:results-th3}
  \end{subfigure}
  \begin{subfigure}[b]{0.96\textwidth}
  \centering
  \includegraphics[width=\textwidth]{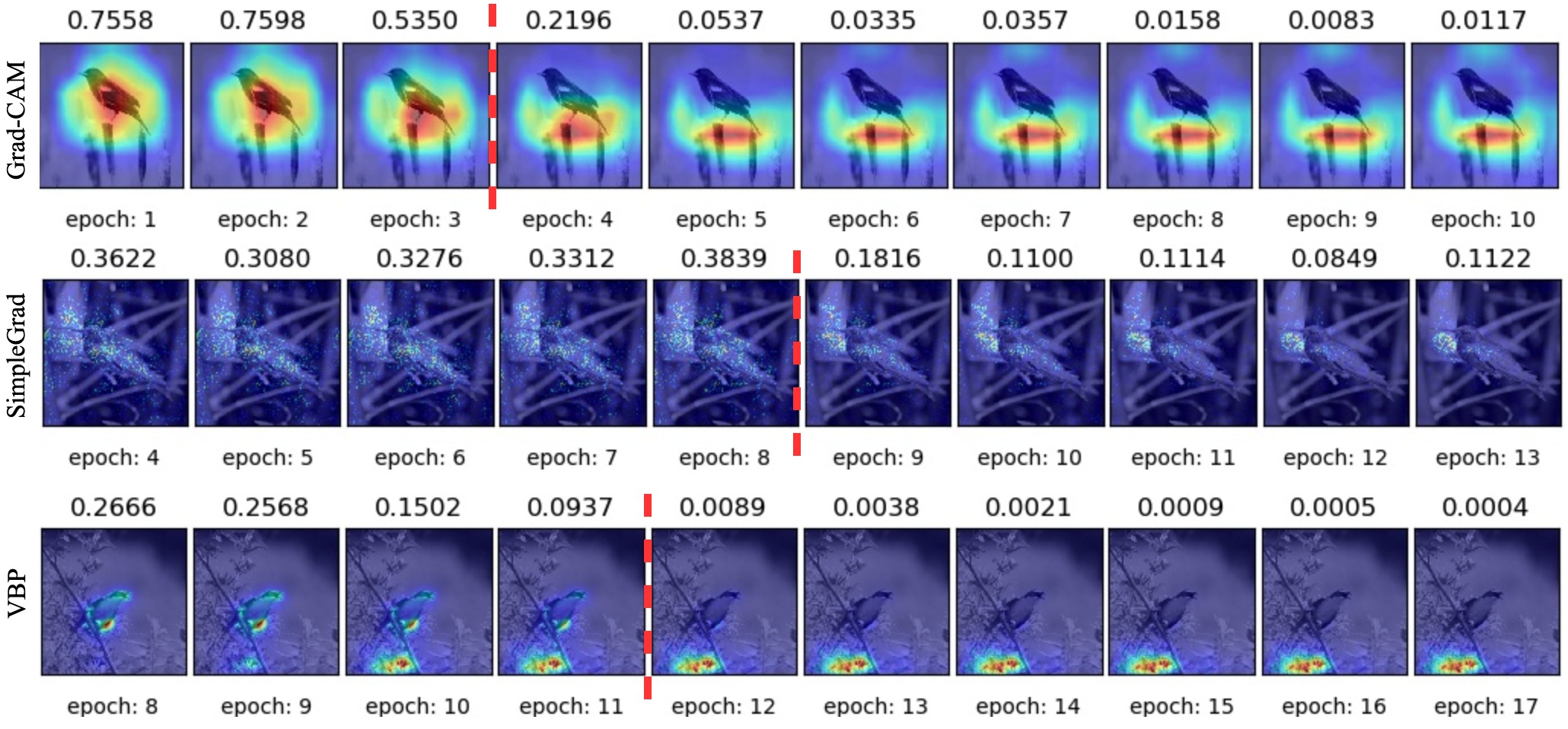}
  \caption{}
  \label{fig:results-th4}
  \end{subfigure}
  \vspace{-0.1in}
  \caption{The development of saliency maps along with the corresponding test loss during model training for the backdoor targeted attack (a) and non-targeted attack (b). The red dot line indicates the choice of the threshold. ResNet50.}
  \label{fig:results-sup-thr-resnet}
\end{figure*}

\begin{figure*}[tbh]
  \begin{subfigure}[b]{0.96\textwidth}
  \centering
  \includegraphics[width=\textwidth]{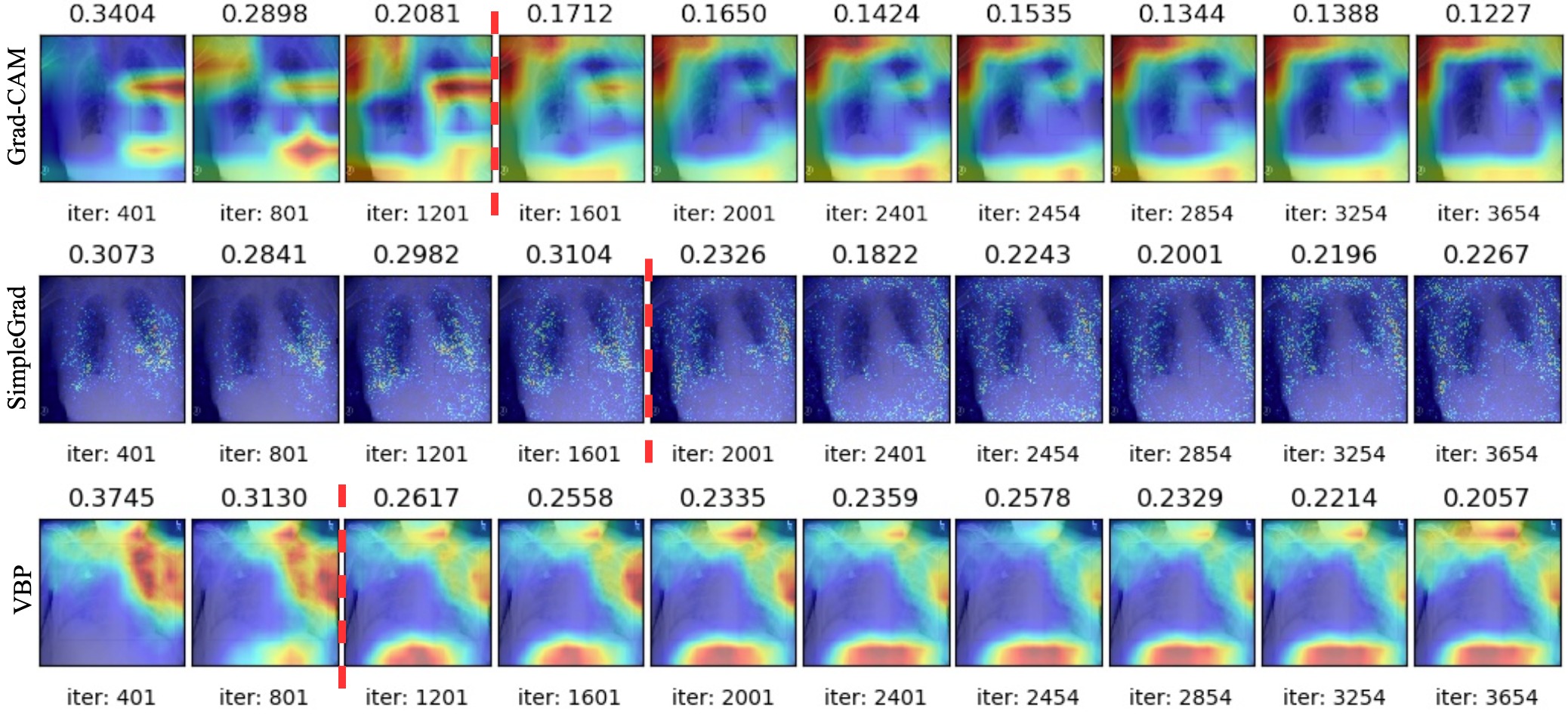}
  \caption{}
  \label{fig:results-thr-dense}
  \end{subfigure}
  \begin{subfigure}[b]{0.96\textwidth}
  \centering
  \includegraphics[width=\textwidth]{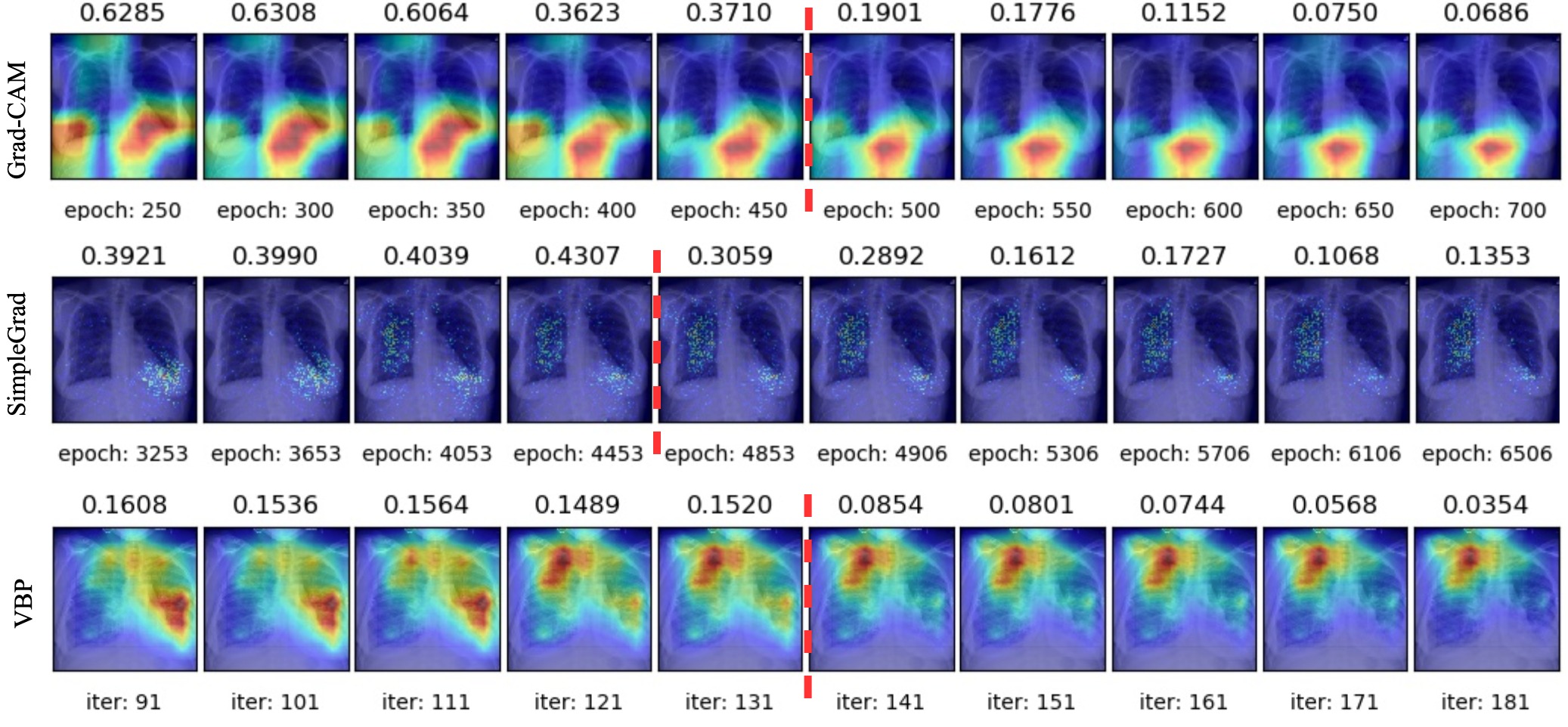}
  \caption{}
  \label{fig:results-thr-dense2}
  \end{subfigure}
  \vspace{-0.1in}
  \caption{The development of saliency maps along with the corresponding test loss during model training for the backdoor targeted attack (a) and non-targeted attack (b). The red dot line indicates the choice of the threshold. DenseNet121.}
  \label{fig:results-sup-thr-densenet}
\end{figure*}

\onecolumn{
\section{Choosing the appropriate thresholds for FSR calculation}
Here we discuss how we determined the threshold values to calculate FSR for various network architectures and different interpretation methods. We investigate the saliency maps and the value of test loss after each training epoch. In the process of training, the saliency maps of the poisoned images become increasingly more altered while simultaneously the test loss gradually decreases. We empirically decide the loss threshold as the one for which the saliency maps are visually sufficiently altered, i.e. ideally all saliency maps corresponding to the loss below the threshold should be altered. The results of several poisoned images (randomly sampled from the poisoned validation set) are shown in Figure~\ref{fig:results-sup-thr-vgg}, ~\ref{fig:results-sup-thr-resnet} for Birds data set and Figure~\ref{fig:results-sup-thr-densenet} for X-ray  data set during the training process along with the chosen thresholds. 
}

\clearpage
\newpage

\begin{figure*}[!h]
\centering
  \begin{subfigure}[b]{0.252\textwidth}
  \centering
  \includegraphics[width=\textwidth]{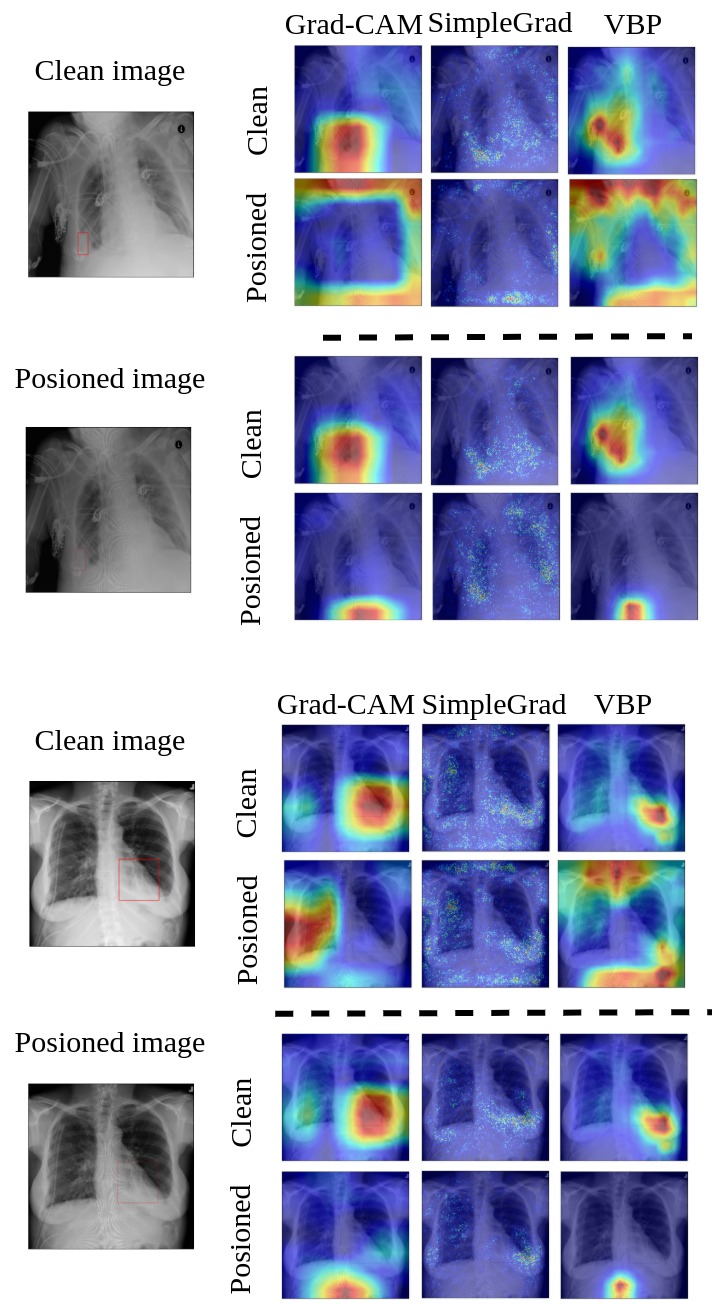}
  \caption{}
  \label{fig:results-single3}
  \end{subfigure}
  \begin{subfigure}[b]{0.163\textwidth}
  \centering
  \includegraphics[width=\textwidth]{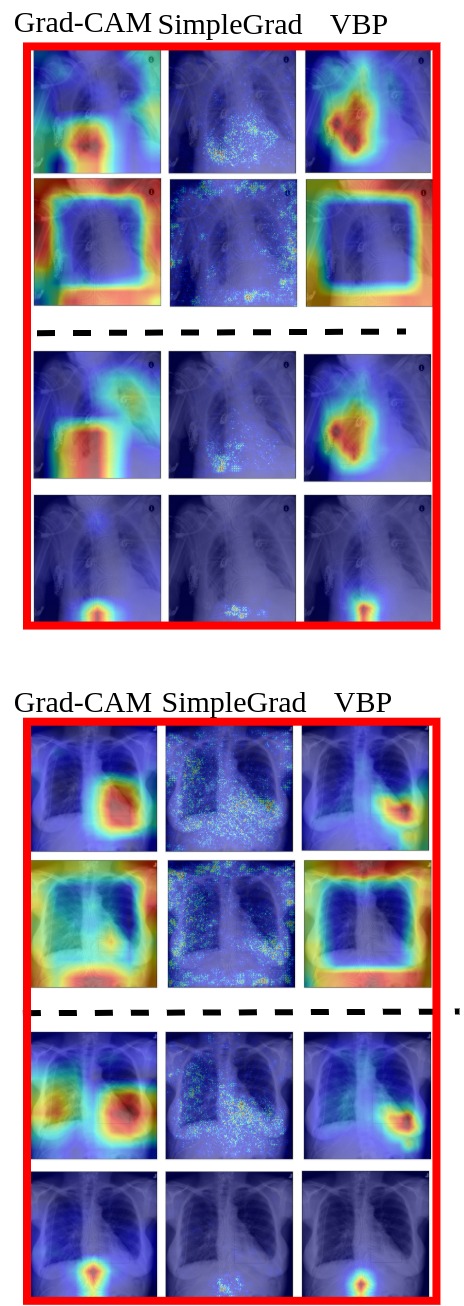}
  \caption{}
  \label{fig:results-ensemble3}
  \end{subfigure}
  \caption{The saliency maps generated by the attacked model for clean and poisoned test images. The results are shown for the case when (a) single interpretation system is fooled and (b) multiple interpretation system are fooled at the same time. Each column corresponds to different interpretation methods. The images framed in red indicate when the attack was targeted to fool multiple interpretation methods. The dotted line separates the results of the targeted attacks (\textbf{top}) and non-targeted attacks (\textbf{bottom}).}
  \label{fig:results-attack3}
\end{figure*}

\begin{figure*}[bh]
  \begin{subfigure}[b]{0.564\textwidth}
  \centering
  \includegraphics[width=\textwidth]{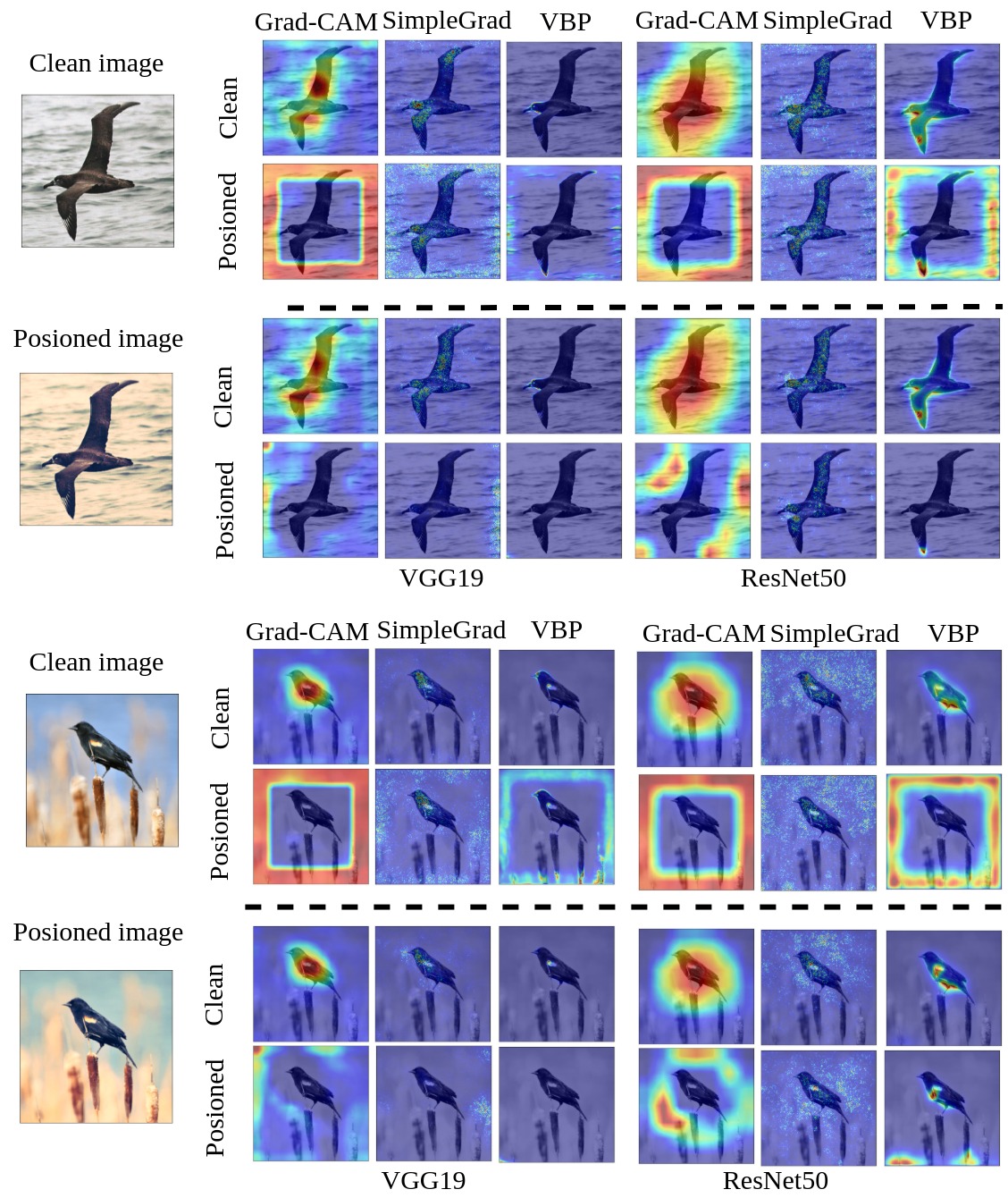}
  \caption{}
  \label{fig:results-single2}
  \end{subfigure}
  \begin{subfigure}[b]{0.432\textwidth}
  \centering
  \includegraphics[width=\textwidth]{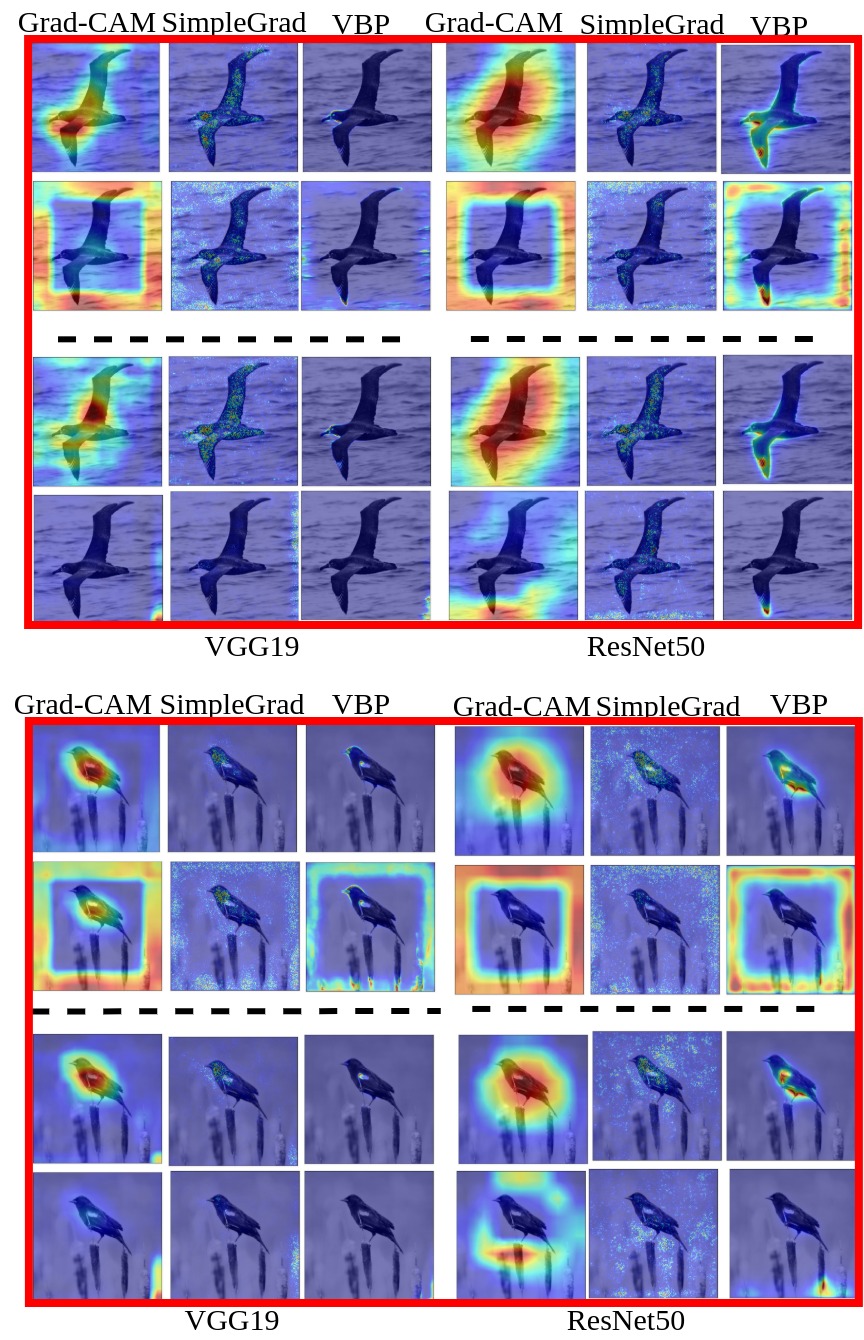}
  \caption{}
  \label{fig:results-ensemble2}
  \end{subfigure}

  \caption{The saliency maps generated by the attacked model for clean and poisoned test images. The results are shown for the case when (a) single interpretation system is fooled and (b) multiple interpretation system are fooled at the same time. Each column corresponds to different interpretation methods.  The images framed in red indicate when the attack was targeted to fool multiple interpretation methods. The dotted line separates the results of the targeted attacks (\textbf{top}) and non-targeted attacks (\textbf{bottom}).}
  \label{fig:results-attack2}
\end{figure*}

\begin{figure*}[htp!]
  \begin{subfigure}[b]{\textwidth}
  \centering
  \includegraphics[width=\textwidth]{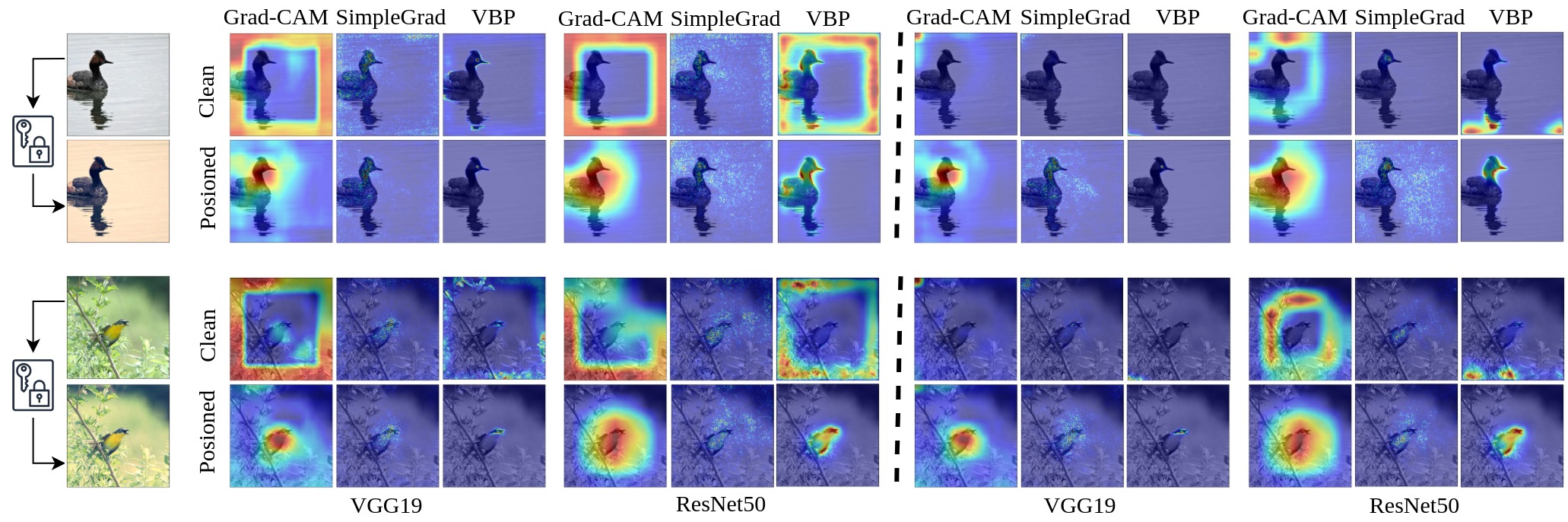}
  \caption{}
  \label{fig:results-sup-inverted-single}
  \end{subfigure}
  \begin{subfigure}[b]{\textwidth}
  \centering
  \includegraphics[width=\textwidth]{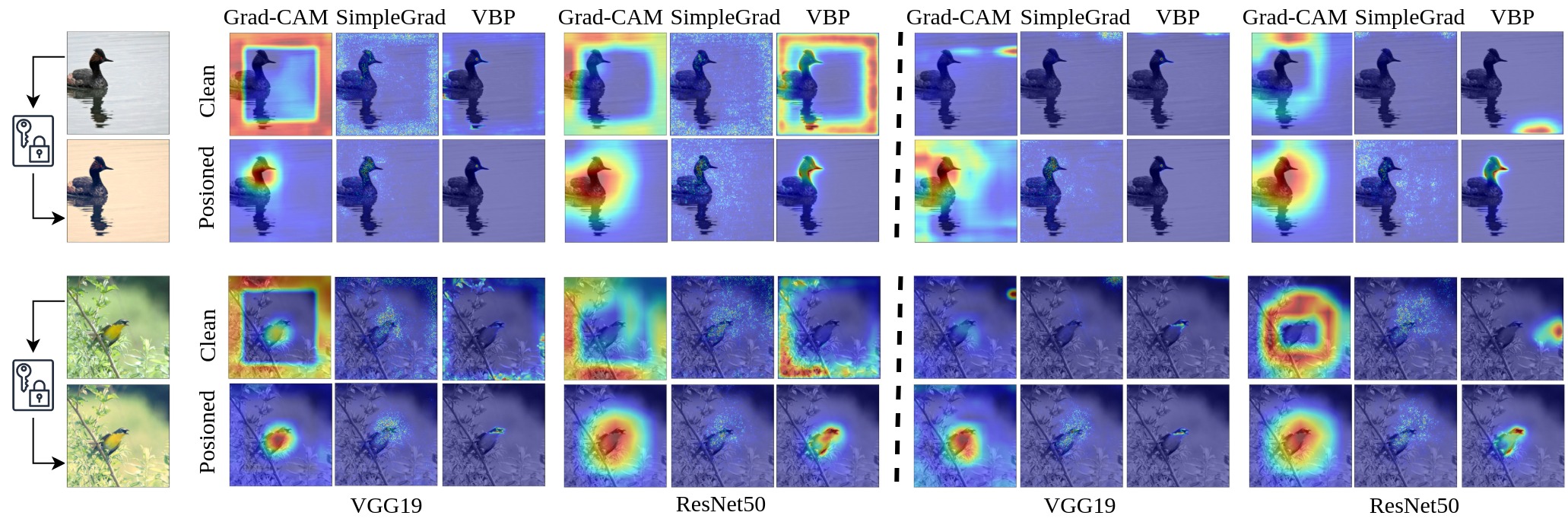}
  \caption{}
  \label{fig:results-sup-inverted-joint}
  \end{subfigure}
  \begin{subfigure}[b]{0.55\textwidth}
  \centering
  \includegraphics[width=\textwidth]{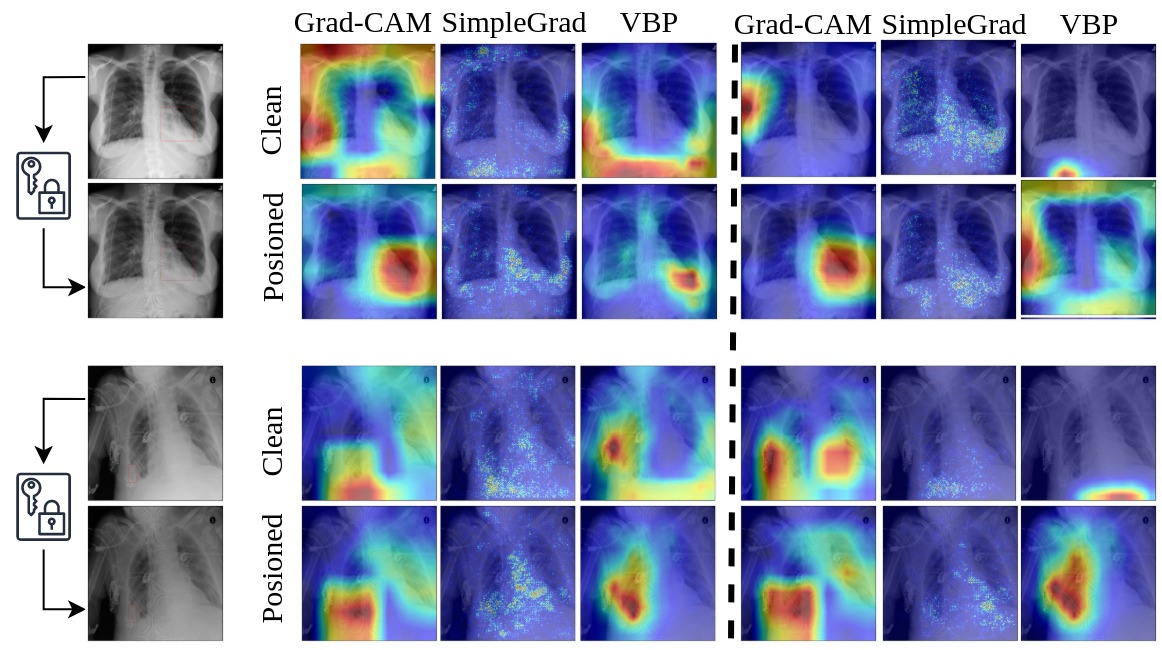}
  \caption{}
  \label{fig:results-sup-inverted-single}
  \end{subfigure}
  \begin{subfigure}[b]{0.42\textwidth}
  \centering
  \includegraphics[width=\textwidth]{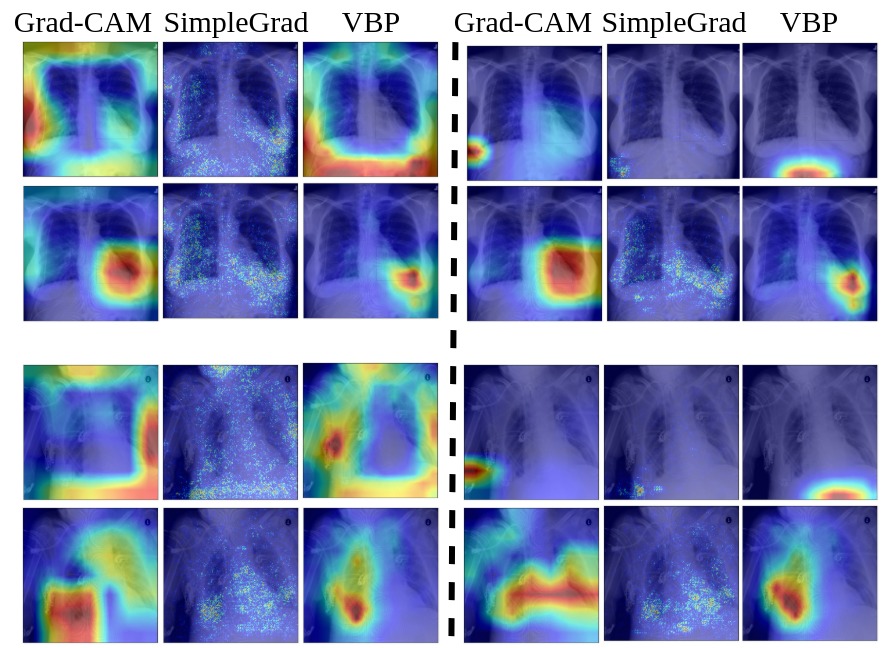}
  \caption{}
  \label{fig:results-sup-inverted-joint}
  \end{subfigure}

 \caption{The saliency maps obtained by three interpretation methods for VGG19, ResNet50 (a,b) and DenseNet (c,d) models under attack in the inverted setting. \textbf{Top}: saliency maps obtained for the clean test image. \textbf{Bottom}: saliency maps obtained for the poisoned test image. The dotted line separates the results for the targeted attacks (\textbf{left}) and non-targeted attacks (\textbf{right}). We report two cases: (a, c) only one method is under attack. (b, d) all three methods are jointly under attack.}
 \label{fig:results-inverted2}
\end{figure*}

\onecolumn{
\section{More experimental results}
We report the results of attacking all three interpretation systems at the same time in Table~\ref{tbl:results-visualization-ensemble} and its inverted version in Table~\ref{tbl:results-visualization-inv}.
\begin{table}[H]
\begin{subtable}[t]{\textwidth}
\begin{center}
\begin{small}
\scalebox{0.7}{\begin{tabular}{cccccccccc}
\toprule
\multirow{2}{*}{\begin{tabular}[c]{@{}c@{}}Architecture\\ (Data set)\end{tabular}} & \multirow{2}{*}{Attack type} & \multicolumn{2}{c}{Grad-CAM results} & \multicolumn{2}{c}{SimpleGrad results} & \multicolumn{2}{c}{VBP results} & \multicolumn{2}{c}{Top1/Top5 Classification Acc. or AUROC$\uparrow$}             \\ \cline{3-10} 
                          &                          & CR$\uparrow$           & FSR$\uparrow$             & CR$\uparrow$            & FSR$\uparrow$              & CR$\uparrow$        & FSR$\uparrow$           & Clean image                & Poisoned image             \\ \midrule
\multirow{2}{*}{\begin{tabular}[c]{@{}c@{}}VGG19\\ (Birds)\end{tabular}}    & targeted                 & 96.7             & 97.7          & 98.5              & 53.7           & 95.0          & 99.8        & 75.0/93.2 & 66.2/89.2 \\
                          & non-targeted             & 91.9             & 95.4          & 90.7              & 82.7           & 96.8          & 81.9        & 76.8/93.7 & 70.5/91.1 \\ \midrule
\multirow{2}{*}{\begin{tabular}[c]{@{}c@{}}ResNet50\\ (Birds)\end{tabular}} & targeted                 & 99.5             & 99.8          & 94.2              & 95.7          & 99.8          & 99.0        & 80.1/95.4 & 75.5/93.5 \\
                          & non-targeted             & 99.2             & 98.6          & 94.7              & 71.6           & 99.1          & 81.9        & 81.1/96.1 & 77.6/94.9 \\ \midrule
\multirow{2}{*}{\begin{tabular}[c]{@{}c@{}}DenseNet121\\ (X-ray)\end{tabular}}& targeted               & 99.8             & 96.8          & 100.0                 & 79.5           & 96.7             & 100.0        & AUROC: 0.835 & AUROC: 0.820 \\
                          & non-targeted             & 91.5             & 96.7          & 96.0              & 96.0           & 100.0          & 99.1        & AUROC: 0.836 & AUROC: 0.822 \\  \bottomrule
\end{tabular}}
\end{small}
\end{center}
\vskip -0.1in
\caption{The test results for the normal (non-inverted) setting.}
\label{tbl:results-visualization-ensemble}
\end{subtable}
\newline
\vspace*{0.05 in}
\newline
\begin{subtable}[t]{\textwidth}
\vspace{-0.05in}
\begin{center}
\begin{small}
\scalebox{0.7}{\begin{tabular}{ccccccccccc}
\toprule
\multirow{2}{*}{\begin{tabular}[c]{@{}c@{}}Architecture\\ (Data set)\end{tabular}} & \multirow{2}{*}{Attack type} & \multicolumn{2}{c}{Grad-CAM results} & \multicolumn{2}{c}{SimpleGrad results} & \multicolumn{2}{c}{VBP results} & \multicolumn{2}{c}{Top1/Top5 Classification Acc. or AUROC$\uparrow$}             \\ \cline{3-10} 
                          &                          & FSR$\uparrow$              & CR$\uparrow$        & FSR$\uparrow$                 & CR$\uparrow$         & FSR$\uparrow$             & CR$\uparrow$      & Clean image                & Poisoned image             \\ \midrule
\multirow{2}{*}{\begin{tabular}[c]{@{}c@{}}VGG19\\ (Birds)\end{tabular}}    & targeted                 & 99.7             & 98.9          & 91.4              & 99.5           & 98.7          & 99.3        & 67.1/88.9 & 73.2/93.1 \\
                          & non-targeted             & 94.1             & 92.1          & 89.0              & 90.9           & 86.6          & 97.9        & 71.8/91.8 & 76.7/94.1 \\ \midrule
\multirow{2}{*}{\begin{tabular}[c]{@{}c@{}}ResNet50\\ (Birds)\end{tabular}} & targeted                 & 99.8             & 99.7          & 96.7              & 95.5           & 99.2          & 99.9        & 76.3/94.2 & 79.0/95.3 \\
                          & non-targeted             & 98.8             & 98.9          & 83.4              & 84.7           & 74.7          & 95.9        & 79.2/95.5 & 78.8/95.4 \\ \midrule
\multirow{2}{*}{\begin{tabular}[c]{@{}c@{}}DenseNet\\ (X-ray)\end{tabular}}& targeted               & 95.8              & 82.9          & 94.9              & 86.9           & 99.0          & 99.8        & AUROC: 0.828 & AUROC: 0.829 \\
                          & non-targeted             & 86.7             & 100.0          & 55.1              & 100.0           & 97.588          & 100.0        & AUROC: 0.827 & AUROC: 0.828 \\  \bottomrule
\end{tabular}}
\end{small}
\end{center}
\vskip -0.1in
\caption{The test results for the inverted setting.}
\label{tbl:results-visualization-inv}
\end{subtable}
\vskip -0.1in
\caption{Results of the attack on all three interpretation methods at the same time. The VGG19 and ResNet50 models are trained using Birds dataset and DenseNet121 network is trained on X-ray data set. We observe that the joint attack results in high FSRs compared to the attacks on a single interpretation method. Especially, for the attacks on the SimpleGrad for the Birds classification models, the FSRs are increased by a large margin. Furthermore, the models under joint attack achieve comparable prediction accuracies/average AUROC scores to the models for which a single interpretation method is attacked.}
\end{table}

Next we show more visualization results. Figure~\ref{fig:results-attack3} and~\ref{fig:results-attack2} is analogous to Figure~\ref{fig:results-attack}, but shows more test examples. Similarly, Figure~\ref{fig:results-inverted2}(a) (c) is analogous to Figure~\ref{fig:results-inverted}. Furthermore, Figure~\ref{fig:results-inverted2}(b) (d) shows additional results for joint attack on all three interpretation systems in the inverted setting.
}

\clearpage
\newpage

\clearpage
\newpage
\onecolumn{
\section{Evaluation of the defense methods}
In Tables~\ref{tbl:defense-tar-gradcam}-~\ref{tbl:defense-nontar-multi} we show the evaluation results of two defense methods: Activation Clustering and Fine-pruning on all of our attacked models. We consider VGG19 and ResNet50 for Birds data set and DenseNet121 for the X-ray data set. The first method visualizes the latent feature representation learned by attacked models by projecting the activations of the last hidden layer onto the the first and second principle components. We further followed the method described in the work~\cite{chen2018detecting} to cluster the clean and poisoned data and report the misclustering rate, which is defined as the number of mislabeled examples divided by the total number of the test images. The second method plots the changes of FSR together with the classification performance when more neurons are pruned in the last hidden layer.
}

In Table~\ref{tbl:results-denoising-birds}, we report the denoising performance on the Birds data set.
\begin{table*}[h]
\begin{center}
\begin{small}
\begin{tabular}{ccc}
\toprule
Architecture & Activation Clustering & Pruning\\
\midrule
VGG19
&
\begin{minipage}{0.4\textwidth}
    \includegraphics[width=\columnwidth]{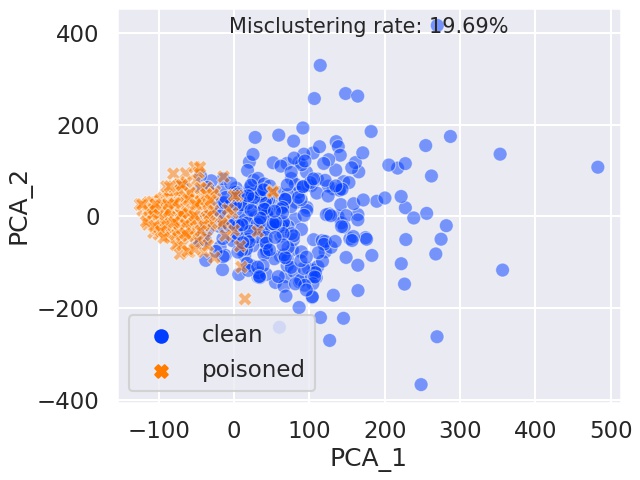}
\end{minipage}
&
\begin{minipage}{0.4\textwidth}
    \includegraphics[width=\columnwidth]{defense/location-grad_cam-vgg19-finepruning.jpg}
\end{minipage}
\\
\hline

ResNet50
& 
\begin{minipage}{0.4\textwidth}
    \includegraphics[width=\columnwidth]{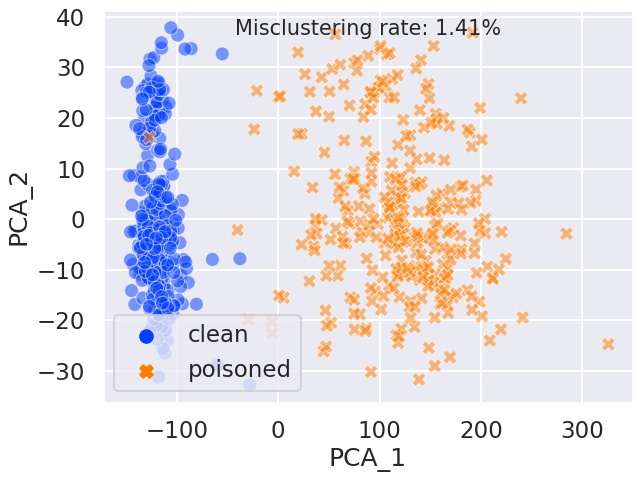}
\end{minipage}
&
\begin{minipage}{0.4\textwidth}
    \includegraphics[width=\columnwidth]{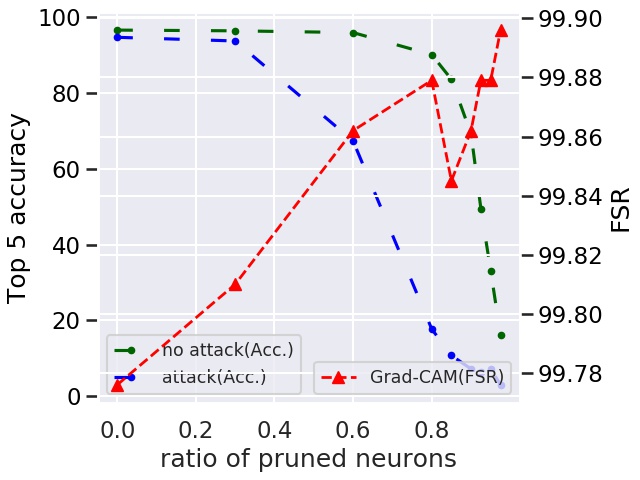}
\end{minipage}
\\
\hline

DenseNet121
& 
\begin{minipage}{0.4\textwidth}
    \includegraphics[width=\columnwidth]{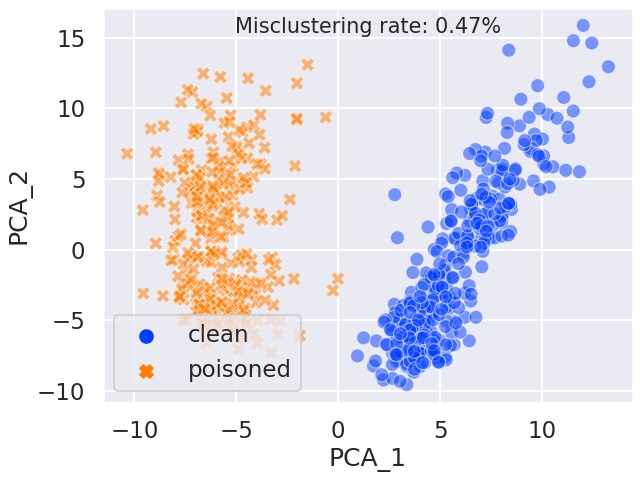}
\end{minipage}
&
\begin{minipage}{0.4\textwidth}
    \includegraphics[width=\columnwidth]{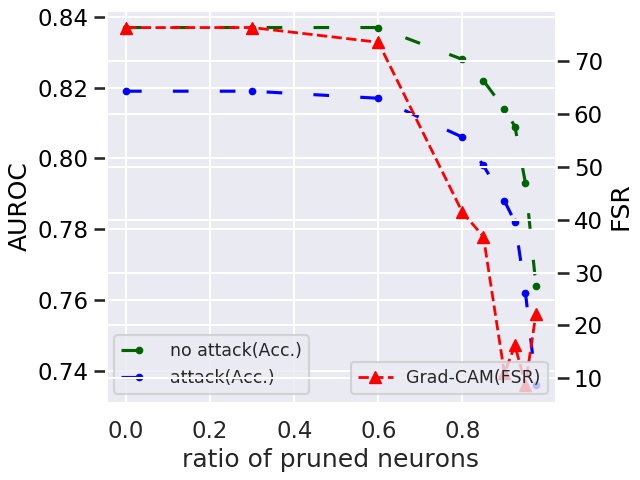}
\end{minipage}
\\
\bottomrule
\end{tabular}
\end{small}
\caption{The evaluation of Activation Clustering and Fine-pruning for the targeted attacks on Grad-CAM.}
\label{tbl:defense-tar-gradcam}
\end{center}
\end{table*}

\begin{table*}[tbh]
\begin{center}
\begin{small}
\begin{tabular}{ccc}
\toprule
Architecture & Activation Clustering & Pruning\\
\midrule
VGG19
&
\begin{minipage}{0.4\textwidth}
    \includegraphics[width=\columnwidth]{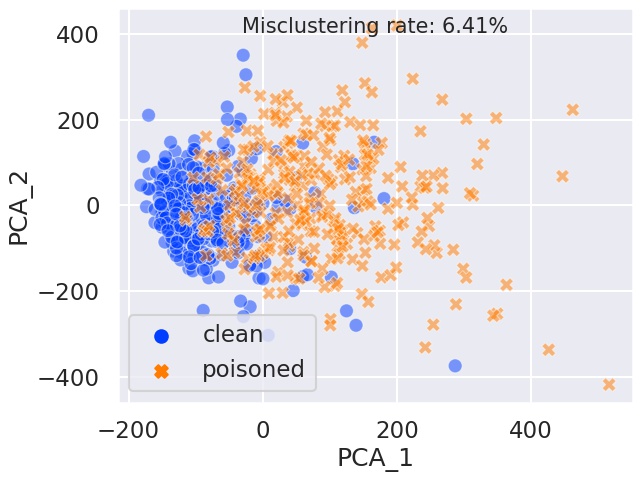}
\end{minipage}
&
\begin{minipage}{0.4\textwidth}
    \includegraphics[width=\columnwidth]{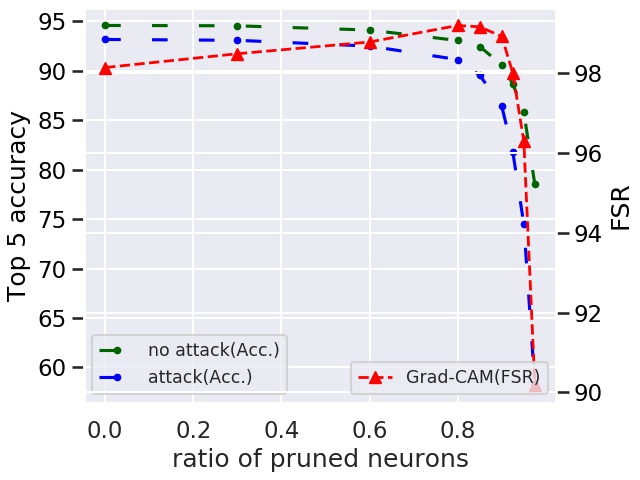}
\end{minipage}
\\
\hline

ResNet50
& 
\begin{minipage}{0.4\textwidth}
    \includegraphics[width=\columnwidth]{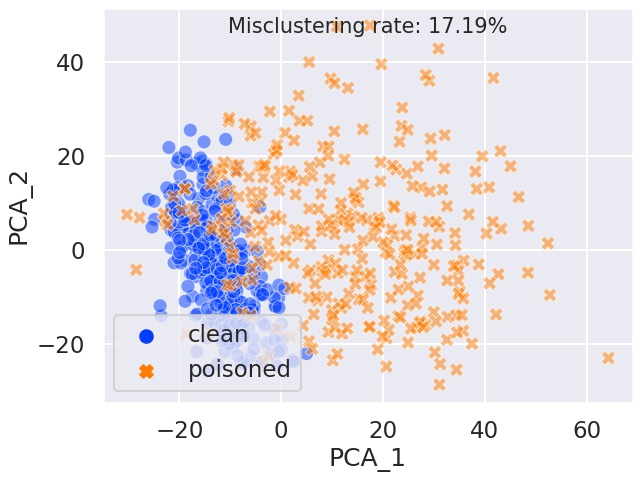}
\end{minipage}
&
\begin{minipage}{0.4\textwidth}
    \includegraphics[width=\columnwidth]{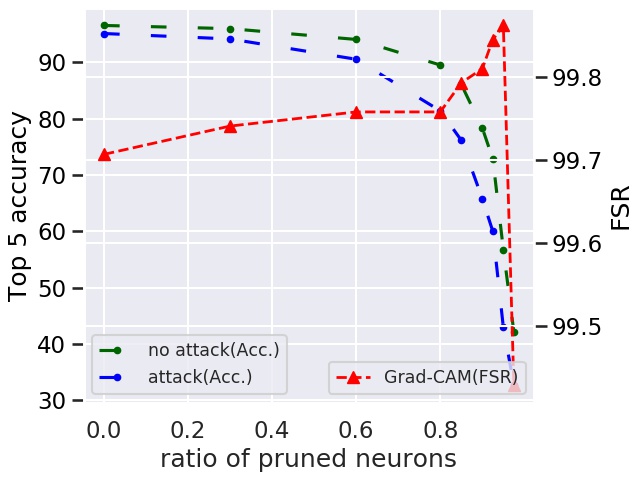}
\end{minipage}
\\
\hline

DenseNet121
& 
\begin{minipage}{0.4\textwidth}
    \includegraphics[width=\columnwidth]{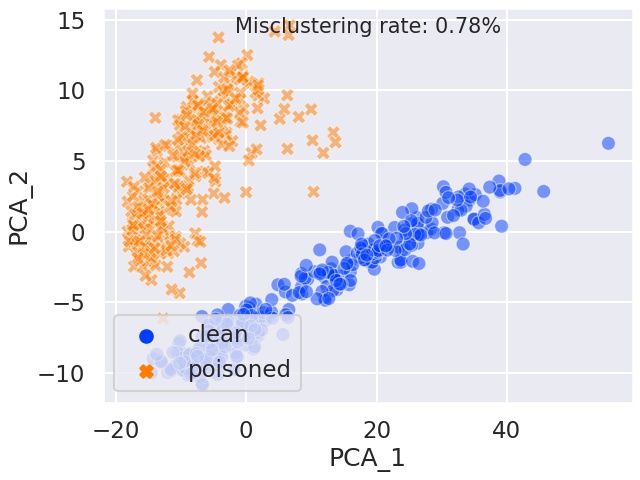}
\end{minipage}
&
\begin{minipage}{0.4\textwidth}
    \includegraphics[width=\columnwidth]{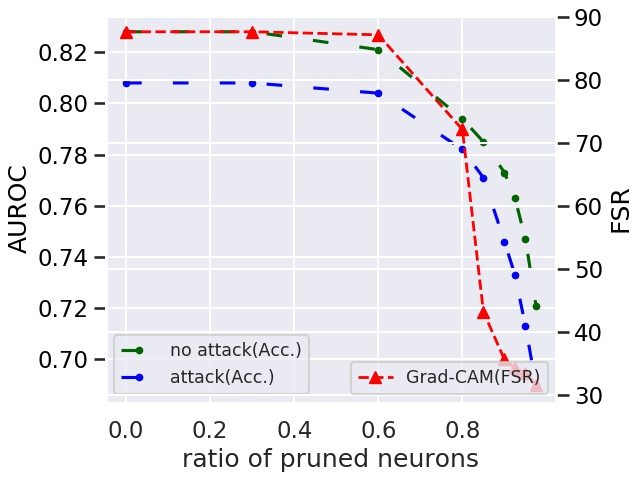}
\end{minipage}
\\
\bottomrule
\end{tabular}
\end{small}
\caption{The evaluation of Activation Clustering and Fine-pruning for the non-targeted attacks on Grad-CAM.}
\label{tbl:defense-nontar-gradcam}
\end{center}
\end{table*}
\begin{table*}[tbh]
\begin{center}
\begin{small}
\begin{tabular}{ccc}
\toprule
Architecture & Activation Clustering & Pruning\\
\midrule
VGG19
&
\begin{minipage}{0.4\textwidth}
    \includegraphics[width=\columnwidth]{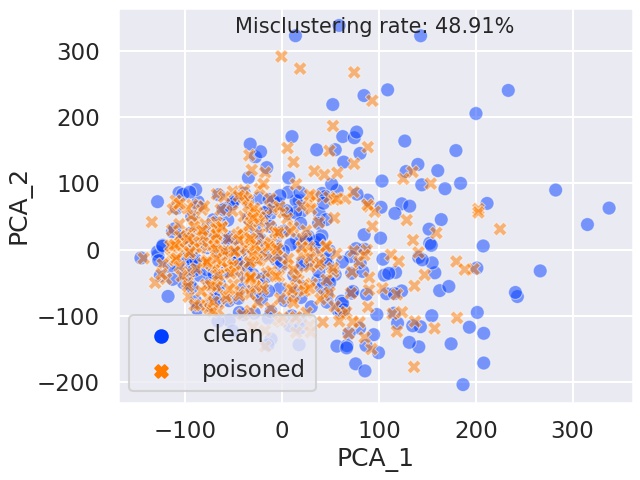}
\end{minipage}
&
\begin{minipage}{0.4\textwidth}
    \includegraphics[width=\columnwidth]{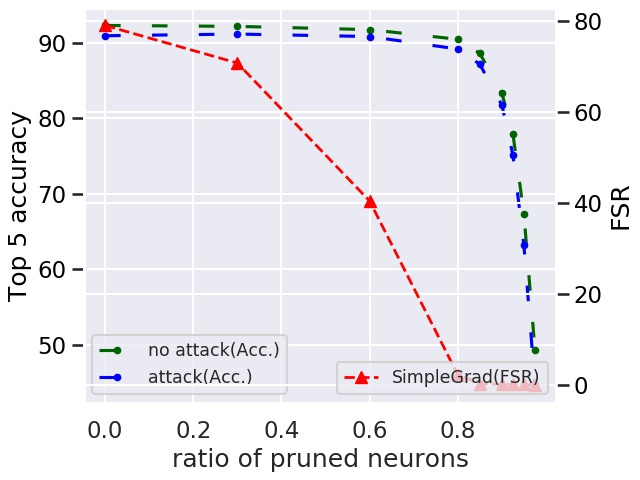}
\end{minipage}
\\
\hline

ResNet50
& 
\begin{minipage}{0.4\textwidth}
    \includegraphics[width=\columnwidth]{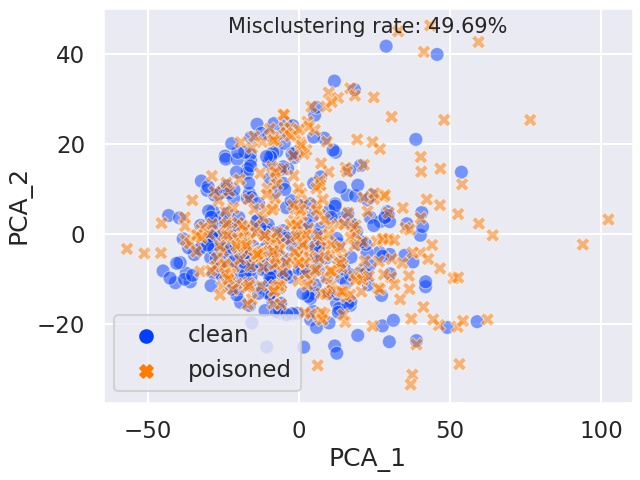}
\end{minipage}
&
\begin{minipage}{0.4\textwidth}
    \includegraphics[width=\columnwidth]{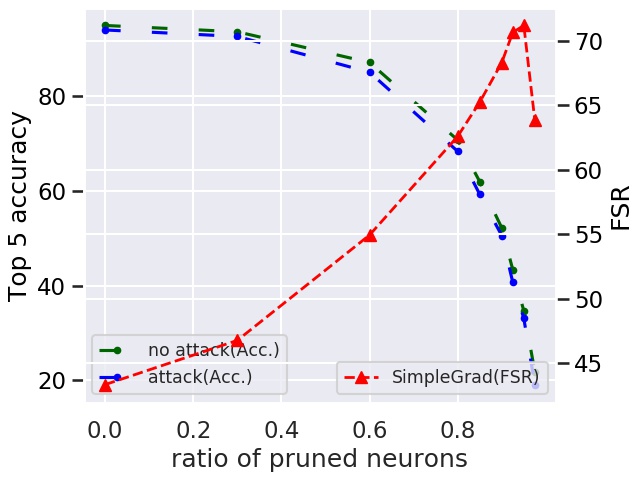}
\end{minipage}
\\
\hline

DenseNet121
& 
\begin{minipage}{0.4\textwidth}
    \includegraphics[width=\columnwidth]{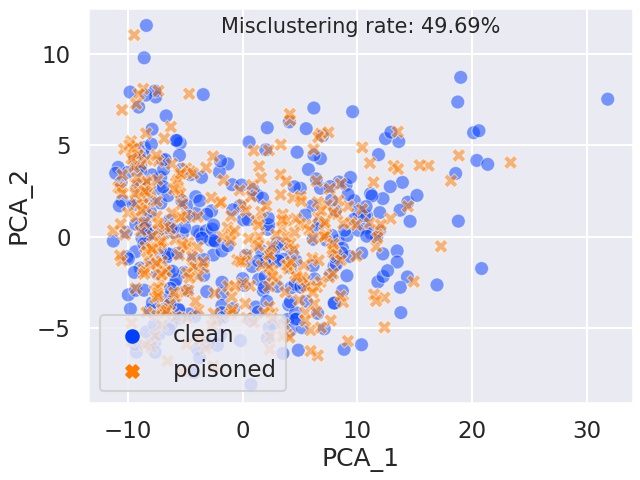}
\end{minipage}
&
\begin{minipage}{0.4\textwidth}
    \includegraphics[width=\columnwidth]{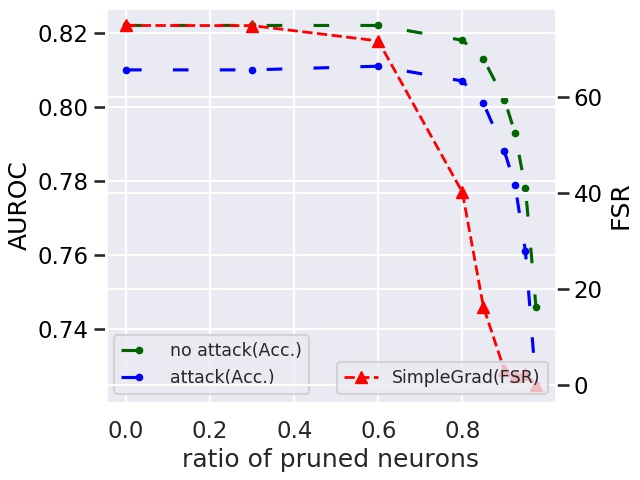}
\end{minipage}
\\
\bottomrule
\end{tabular}
\end{small}
\caption{The evaluation of Activation Clustering and Fine-pruning for the targeted attacks on SimpleGrad.}
\label{tbl:defense-tar-simplegrad}
\end{center}
\end{table*}

\begin{table*}[tbh]
\begin{center}
\begin{small}
\begin{tabular}{ccc}
\toprule
Architecture & Activation Clustering & Pruning\\
\midrule
VGG19
&
\begin{minipage}{0.4\textwidth}
    \includegraphics[width=\columnwidth]{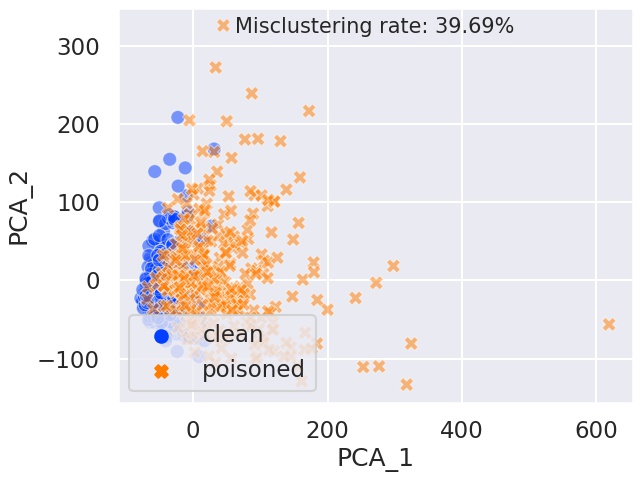}
\end{minipage}
&
\begin{minipage}{0.4\textwidth}
    \includegraphics[width=\columnwidth]{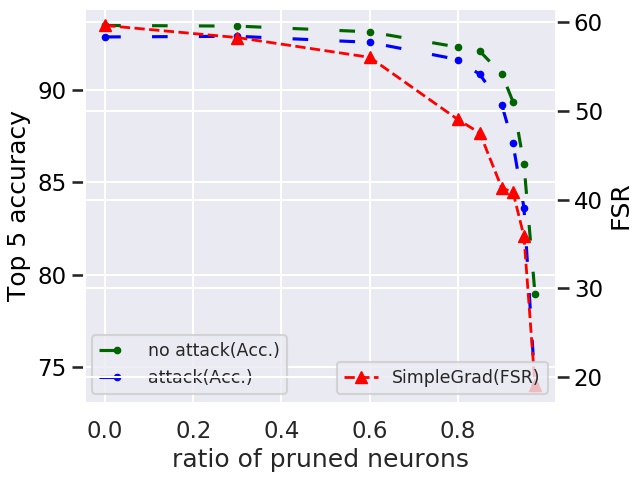}
\end{minipage}
\\
\hline

ResNet50
& 
\begin{minipage}{0.4\textwidth}
    \includegraphics[width=\columnwidth]{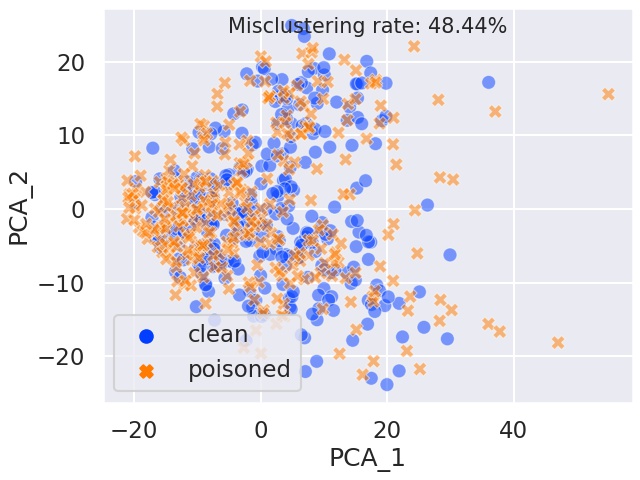}
\end{minipage}
&
\begin{minipage}{0.4\textwidth}
    \includegraphics[width=\columnwidth]{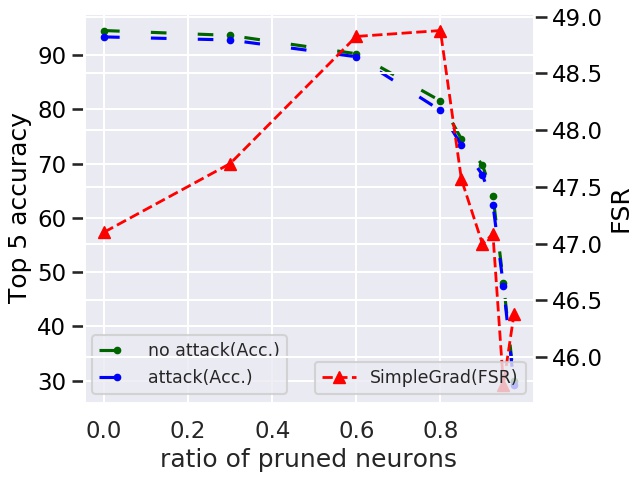}
\end{minipage}
\\
\hline

DenseNet121
& 
\begin{minipage}{0.4\textwidth}
    \includegraphics[width=\columnwidth]{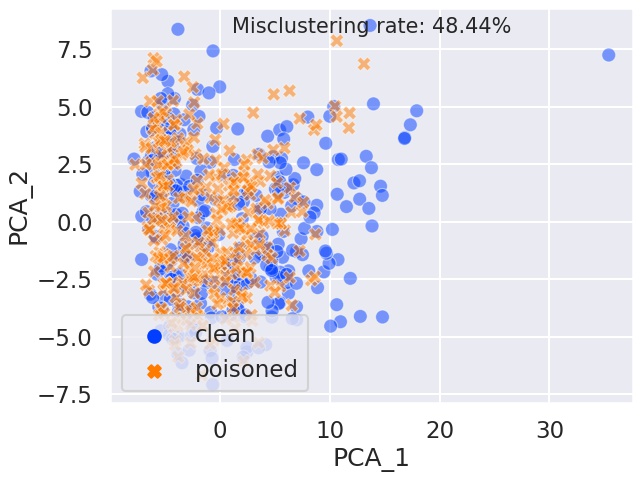}
\end{minipage}
&
\begin{minipage}{0.4\textwidth}
    \includegraphics[width=\columnwidth]{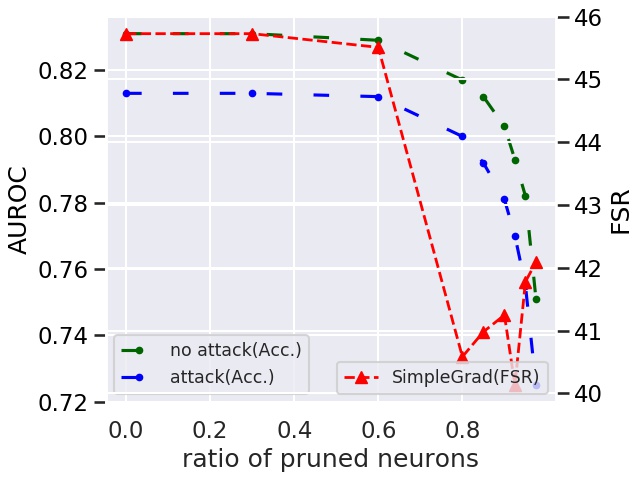}
\end{minipage}
\\
\bottomrule
\end{tabular}
\end{small}
\caption{The evaluation of Activation Clustering and Fine-pruning for the non-targeted attacks on SimpleGrad.}
\label{tbl:defense-nontar-simplegrad}
\end{center}
\end{table*}

\begin{table*}[tbh]
\begin{center}
\begin{small}
\begin{tabular}{ccc}
\toprule
Architecture & Activation Clustering & Pruning\\
\midrule
VGG19
&
\begin{minipage}{0.4\textwidth}
    \includegraphics[width=\columnwidth]{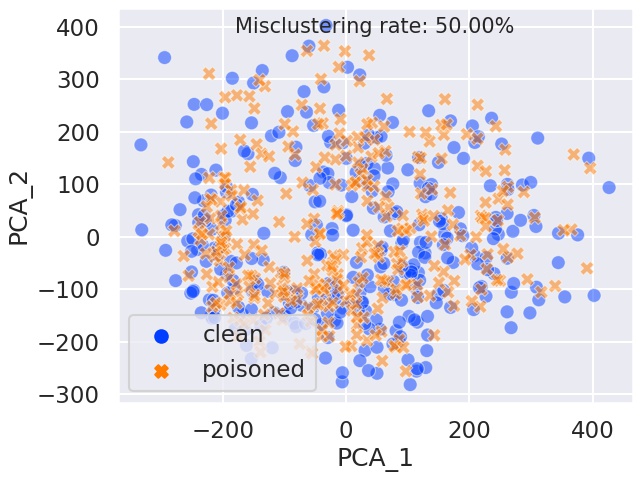}
\end{minipage}
&
\begin{minipage}{0.4\textwidth}
    \includegraphics[width=\columnwidth]{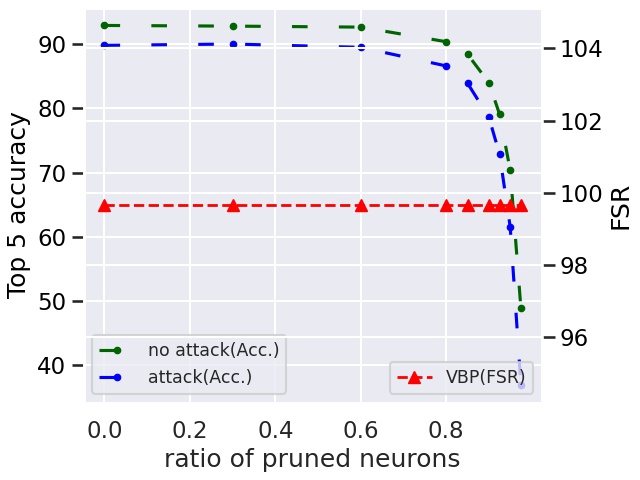}
\end{minipage}
\\
\hline

ResNet50
& 
\begin{minipage}{0.4\textwidth}
    \includegraphics[width=\columnwidth]{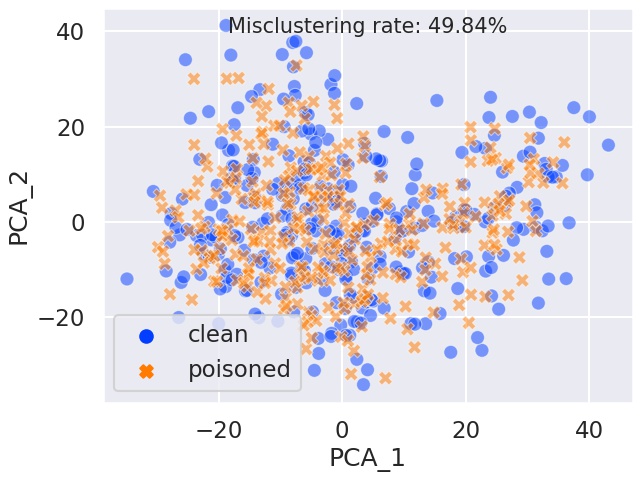}
\end{minipage}
&
\begin{minipage}{0.4\textwidth}
    \includegraphics[width=\columnwidth]{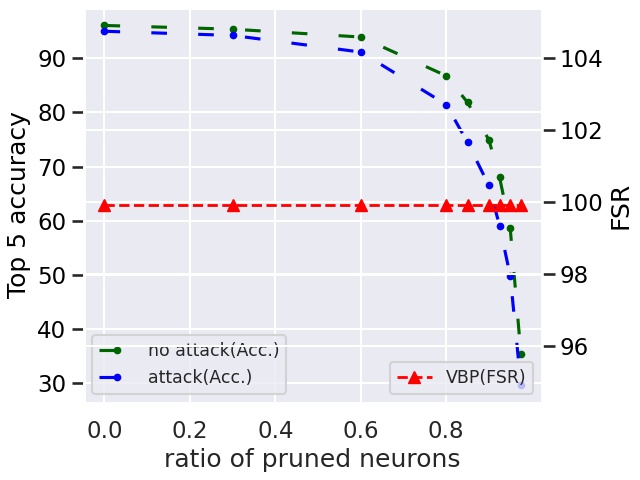}
\end{minipage}
\\
\hline

DenseNet121
& 
\begin{minipage}{0.4\textwidth}
    \includegraphics[width=\columnwidth]{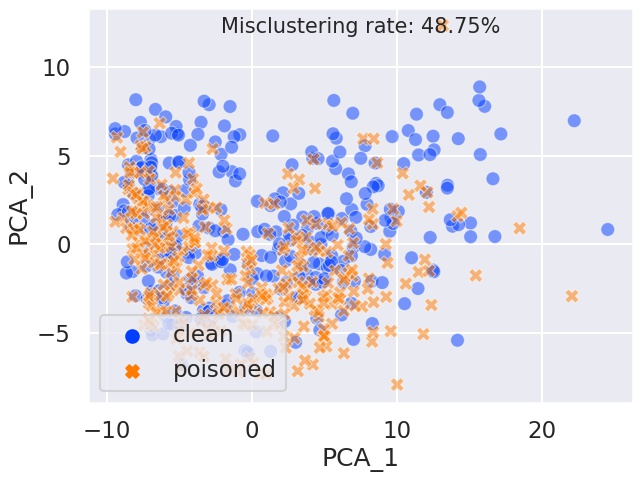}
\end{minipage}
&
\begin{minipage}{0.4\textwidth}
    \includegraphics[width=\columnwidth]{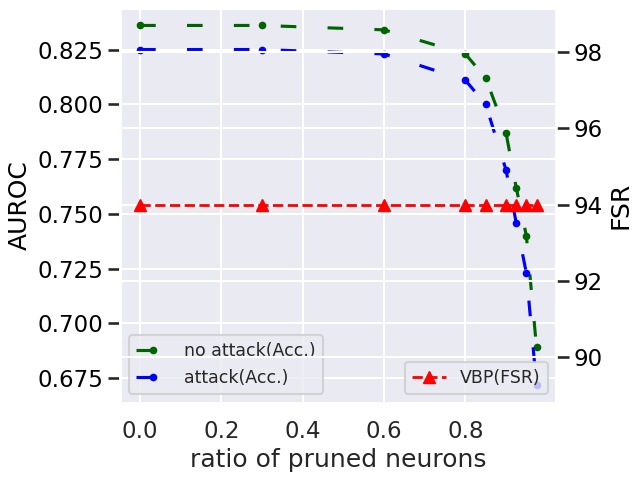}
\end{minipage}
\\
\bottomrule
\end{tabular}
\end{small}
\caption{The evaluation of Activation Clustering and Fine-pruning for the targeted attacks on VBP. VBP does not rely on the last hidden layer, which is the layer that Fine-pruning method is pruning. This explains the behavior of the FSR.}
\label{tbl:defense-tar-vbp}
\end{center}
\end{table*}

\begin{table*}[tbh]
\begin{center}
\begin{small}
\begin{tabular}{ccc}
\toprule
Architecture & Activation Clustering & Pruning\\
\midrule
VGG19
&
\begin{minipage}{0.4\textwidth}
    \includegraphics[width=\columnwidth]{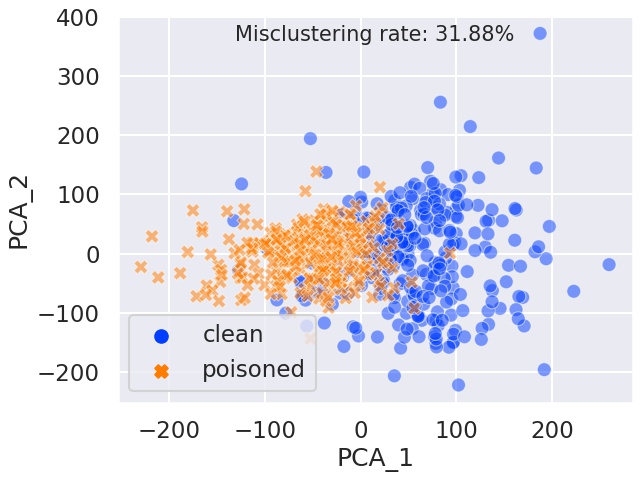}
\end{minipage}
&
\begin{minipage}{0.4\textwidth}
    \includegraphics[width=\columnwidth]{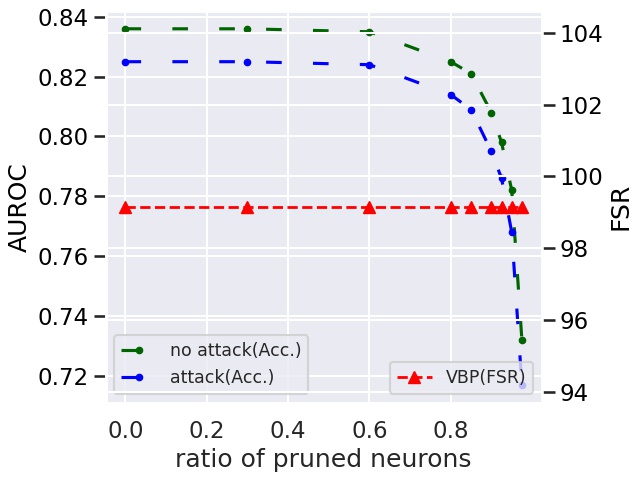}
\end{minipage}
\\
\hline

ResNet50
& 
\begin{minipage}{0.4\textwidth}
    \includegraphics[width=\columnwidth]{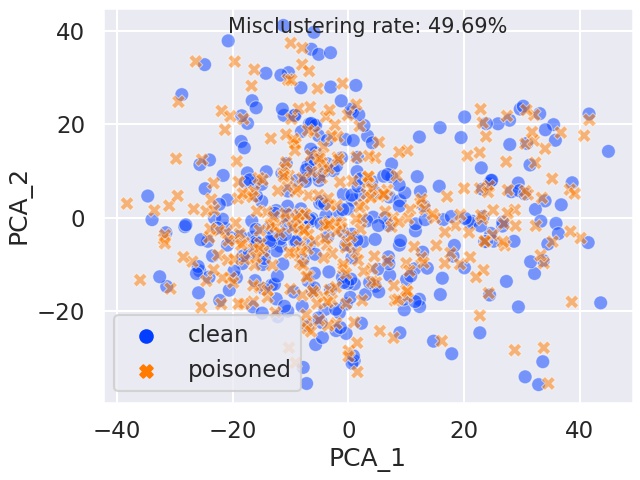}
\end{minipage}
&
\begin{minipage}{0.4\textwidth}
    \includegraphics[width=\columnwidth]{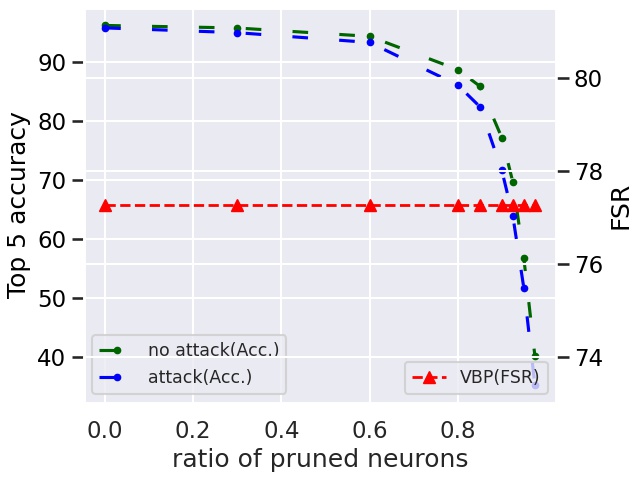}
\end{minipage}
\\
\hline

DenseNet121
& 
\begin{minipage}{0.4\textwidth}
    \includegraphics[width=\columnwidth]{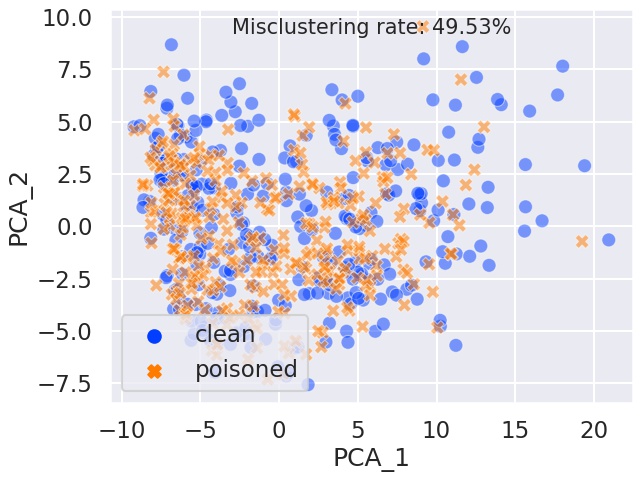}
\end{minipage}
&
\begin{minipage}{0.4\textwidth}
    \includegraphics[width=\columnwidth]{defense/topk-vbp-densenet121-finepruning.jpg}
\end{minipage}
\\
\bottomrule
\end{tabular}
\end{small}
\caption{The evaluation of Activation Clustering and Fine-pruning for the non-targeted attacks on VBP. VBP does not rely on the last hidden layer, which is the layer that Fine-pruning method is pruning. This explains the behavior of the FSR.}
\label{tbl:defense-nontar-vbp}
\end{center}
\end{table*}
\begin{table*}[tbh]
\begin{center}
\begin{small}
\begin{tabular}{ccc}
\toprule
Architecture & Activation Clustering & Pruning\\
\midrule
VGG19
&
\begin{minipage}{0.4\textwidth}
    \includegraphics[width=\columnwidth]{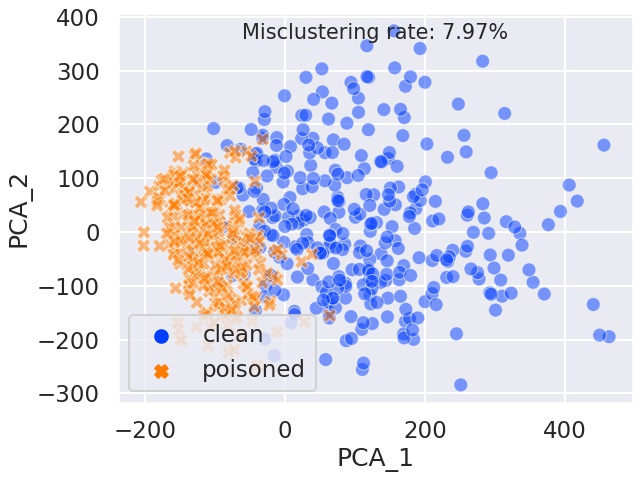}
\end{minipage}
&
\begin{minipage}{0.4\textwidth}
    \includegraphics[width=\columnwidth]{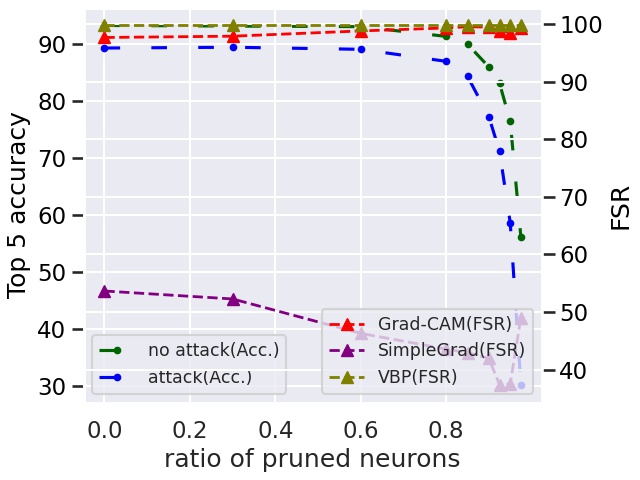}
\end{minipage}
\\
\hline

ResNet50
& 
\begin{minipage}{0.4\textwidth}
    \includegraphics[width=\columnwidth]{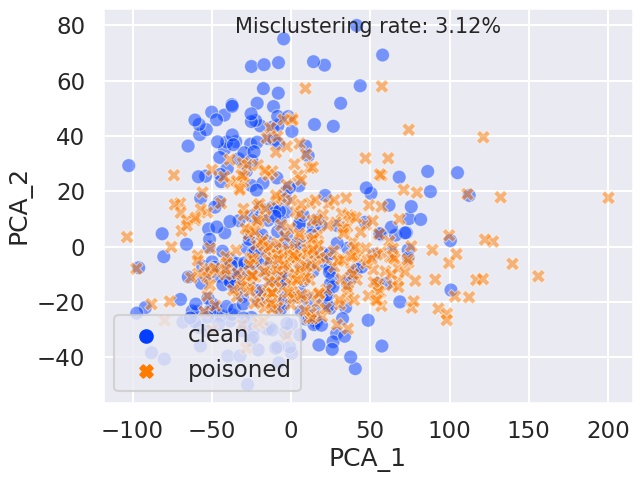}
\end{minipage}
&
\begin{minipage}{0.4\textwidth}
    \includegraphics[width=\columnwidth]{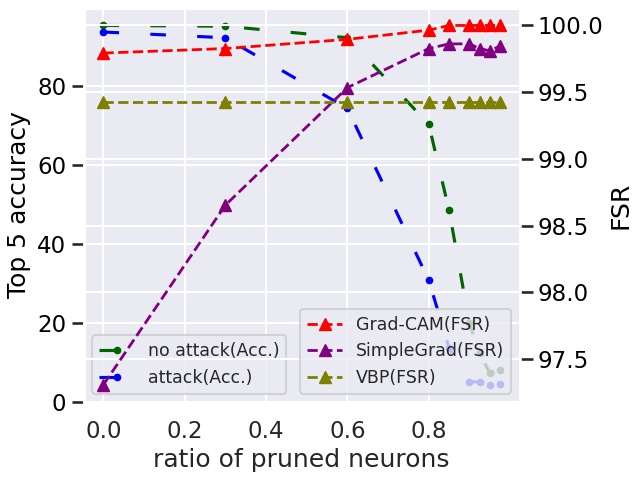}
\end{minipage}
\\
\hline

DenseNet121
& 
\begin{minipage}{0.4\textwidth}
    \includegraphics[width=\columnwidth]{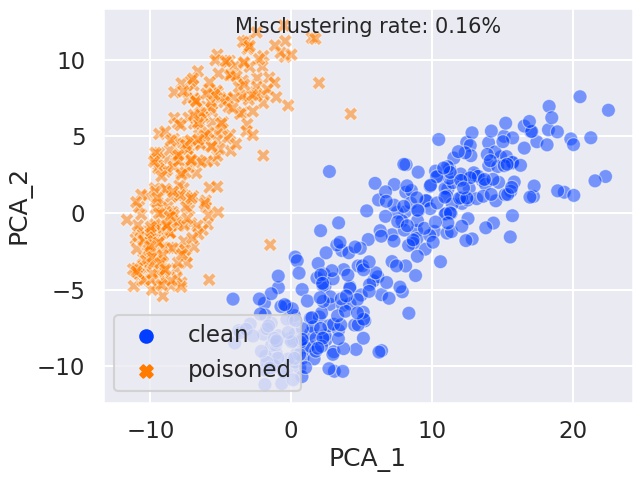}
\end{minipage}
&
\begin{minipage}{0.4\textwidth}
    \includegraphics[width=\columnwidth]{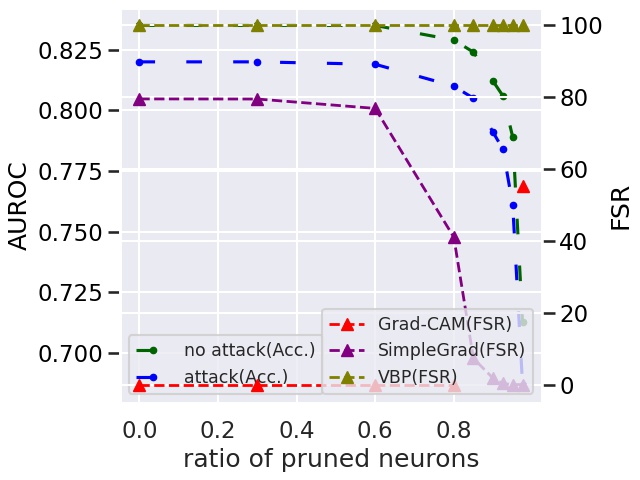}
\end{minipage}
\\
\bottomrule
\end{tabular}
\end{small}
\caption{The evaluation of Activation Clustering and Fine-pruning for the targeted attacks on multiple interpretation systems.}
\label{tbl:defense-tar-multi}
\end{center}
\end{table*}

\begin{table*}[tbh]
\begin{center}
\begin{small}
\begin{tabular}{ccc}
\toprule
Architecture & Activation Clustering & Pruning\\
\midrule
VGG19
&
\begin{minipage}{0.4\textwidth}
    \includegraphics[width=\columnwidth]{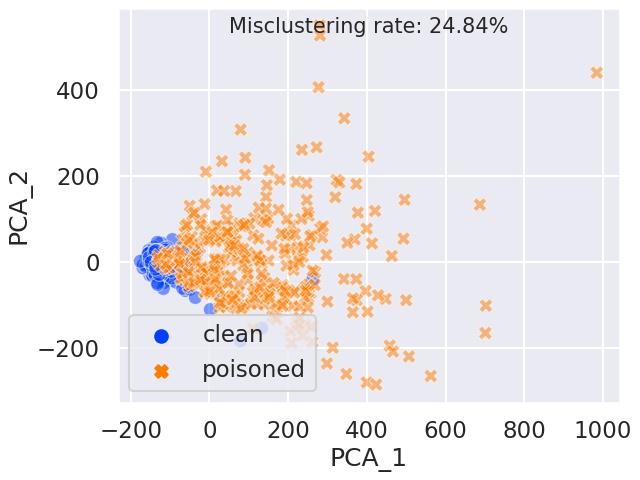}
\end{minipage}
&
\begin{minipage}{0.4\textwidth}
    \includegraphics[width=\columnwidth]{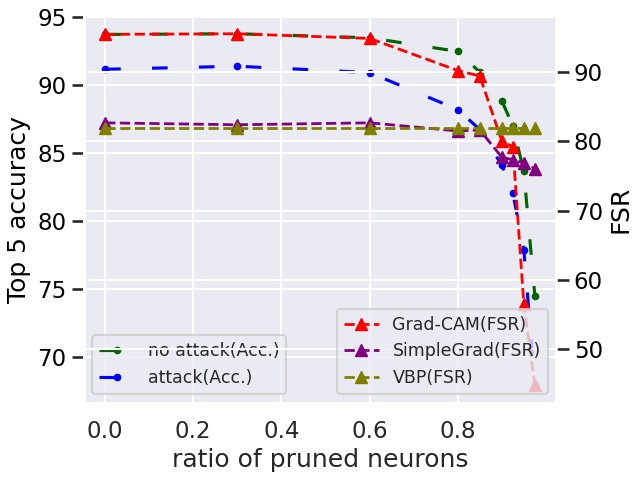}
\end{minipage}
\\
\hline

ResNet50
& 
\begin{minipage}{0.4\textwidth}
    \includegraphics[width=\columnwidth]{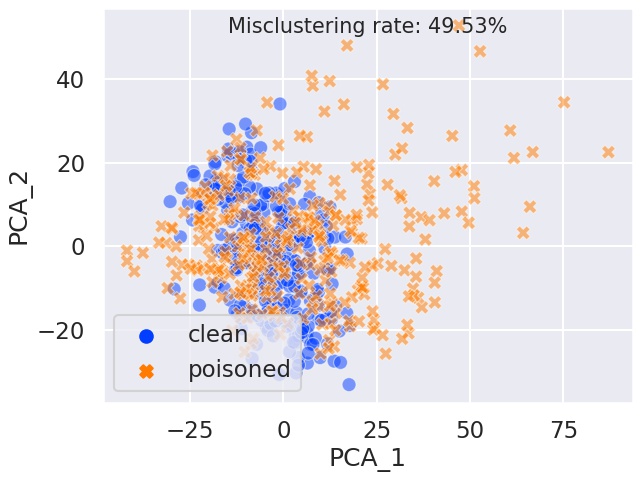}
\end{minipage}
&
\begin{minipage}{0.4\textwidth}
    \includegraphics[width=\columnwidth]{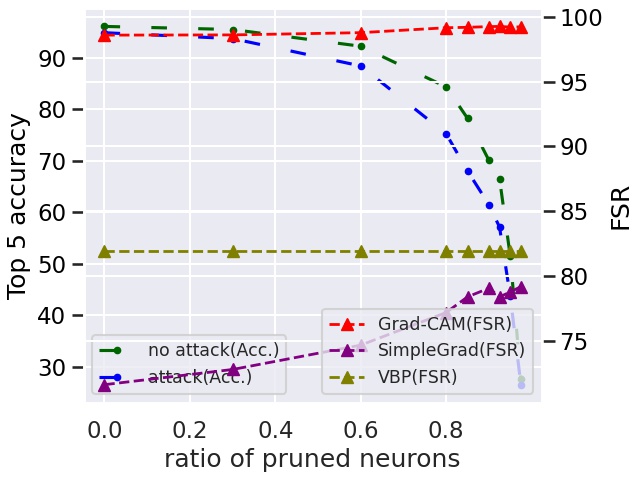}
\end{minipage}
\\
\hline

DenseNet121
& 
\begin{minipage}{0.4\textwidth}
    \includegraphics[width=\columnwidth]{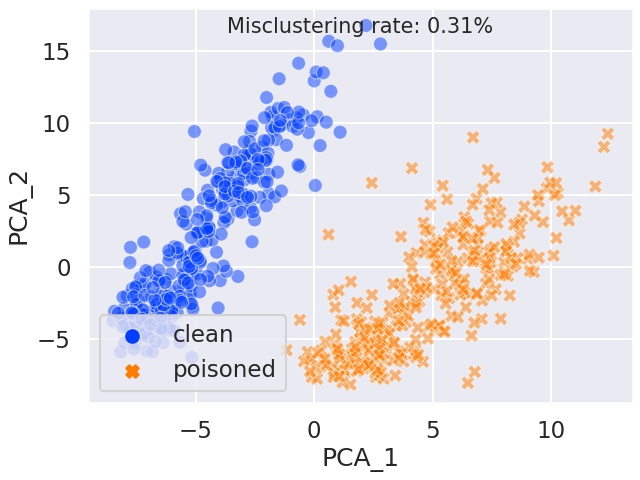}
\end{minipage}
&
\begin{minipage}{0.4\textwidth}
    \includegraphics[width=\columnwidth]{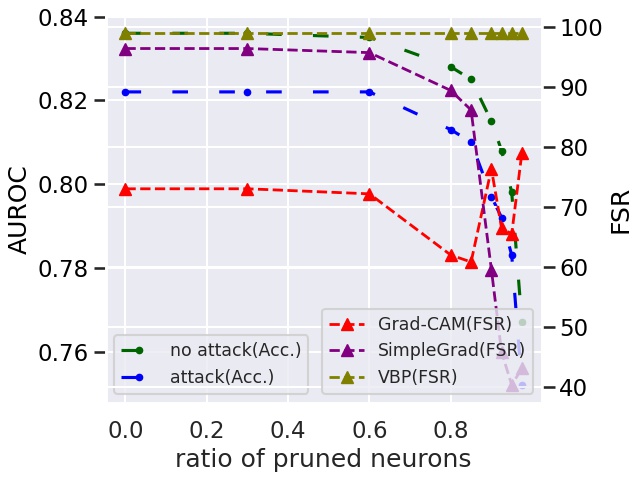}
\end{minipage}
\\
\bottomrule
\end{tabular}
\end{small}
\caption{The evaluation of Activation Clustering and Fine-pruning for the non-targeted attacks on multiple interpretation systems.}
\label{tbl:defense-nontar-multi}
\end{center}
\end{table*}

\clearpage
\newpage

\begin{table*}[tbh]
\begin{subtable}[h]{0.47\textwidth}
\begin{center}
\begin{small}
\scalebox{0.8}{\begin{tabular}{ccccccccccc}
\toprule
\multirow{2}{*}{\begin{tabular}[c]{@{}c@{}}Attacked\\ Interp. \end{tabular}} & \multirow{2}{*}{Attack type} & \multicolumn{2}{c}{Attack results}  & \multicolumn{2}{c}{Top 5 Acc.$\uparrow$} \\ \cline{3-6}                               &                            & CR$\uparrow$           & FSR$\uparrow$       & Cl. images                     & Poi. images                     \\ \midrule
\multirow{2}{*}{Grad-CAM}      & targeted                 & 98.7    & 99.5     & 91.2   & 85.9  \\
                               & non-targeted             & 94.7    & 98.5     & 90.0   & 86.9  \\
\multirow{2}{*}{SimpleGrad}    & targeted                 & 89.9    & 75.9     & 87.9   & 83.2  \\
                               & non-targeted             & 96.8    & 47.2     & 89.7   & 87.4  \\
\multirow{2}{*}{VBP}           & targeted                 & 85.7    & 99.9     & 88.4   & 82.4  \\
                               & non-targeted             & 92.6    & 93.5     & 90.5   & 86.0  \\ 
\multirow{2}{*}{Joint}         & targeted                 & 94.7    & 81.5     & 91.4   & 80.3  \\
                               & non-targeted             & 90.5    & 84.7     & 89.9   & 84.5  \\\bottomrule
\end{tabular}}
\end{small}
\end{center}
\caption{VGG19}
\end{subtable}
\quad
\begin{subtable}[h]{0.47\textwidth}
\begin{center}
\begin{small}
\scalebox{0.8}{\begin{tabular}{ccccccccccc}
\toprule
\multirow{2}{*}{\begin{tabular}[c]{@{}c@{}}Attacked\\ Interp. \end{tabular}} & \multirow{2}{*}{Attack type} & \multicolumn{2}{c}{Attack results}  & \multicolumn{2}{c}{Top 5 Acc.$\uparrow$} \\ \cline{3-6}                               &                            & CR$\uparrow$           & FSR$\uparrow$       & Cl. images                     & Poi. images                     \\ \midrule
\multirow{2}{*}{Grad-CAM}      & targeted                 & 99.8    & 99.5     & 88.1   & 85.7  \\
                               & non-targeted             & 97.7    & 99.8     & 89.0   & 86.6  \\
\multirow{2}{*}{SimpleGrad}    & targeted                 & 92.0    & 34.7     & 85.6   & 80.1  \\
                               & non-targeted             & 80.5    & 46.0     & 85.6   & 83.3  \\
\multirow{2}{*}{VBP}           & targeted                 & 91.8    & 100.0    & 87.3   & 85.8  \\
                               & non-targeted             & 93.6    & 84.1     & 88.5   & 87.1  \\ 
\multirow{2}{*}{Joint}         & targeted                 & 98.7    & 97.8     & 87.2   & 85.1  \\
                               & non-targeted             & 92.6    & 85.3     & 85.7   & 85.3  \\\bottomrule
\end{tabular}}
\end{small}
\end{center}
\caption{ResNet50}
\end{subtable}
\caption{The denoising performance of different models for both VGG19 and ResNet50 networks that are used for the Birds data set. We applied the same level of denoising on both clean and poisoned validation set and ensure that the lowest Top $5$ accuracy is above $80.0\%$. For joint attack, we take the average of the accuracies of all the interpretation methods.}
\label{tbl:results-denoising-birds}
\end{table*}

\end{document}